\documentclass[twocolumn,showpacs,amsmath,amssymb,prd]{revtex4}
\def\ba{\begin{eqnarray}}
\def\ea{\end{eqnarray}}

\usepackage{graphicx}
\usepackage{dcolumn}
\usepackage{bm}
\usepackage{epsf}

\begin{document}

\title{Relativistic Models for Binary Neutron Stars with Arbitrary Spins}

\author{Pedro Marronetti}
\altaffiliation{Fortner Fellow}

\affiliation{Department of Physics, University of Illinois at Urbana-Champaign, Urbana, IL 61801}

\author{Stuart L. Shapiro}
\altaffiliation{Department of Astronomy \& NCSA, University of Illinois at Urbana-Champaign, Urbana, IL 61801}

\affiliation{Department of Physics, University of Illinois at Urbana-Champaign, Urbana, IL 61801}

\begin{abstract}
We introduce a new numerical scheme for solving the initial value problem for quasiequilibrium binary neutron stars allowing for arbitrary spins. The coupled Einstein field equations and equations of relativistic hydrodynamics are solved in the Wilson-Mathews conformal thin sandwich formalism. We construct sequences of 
circular-orbit binaries of varying separation, keeping the rest mass and circulation constant along each sequence. Solutions are presented for configurations obeying an $n=1$ polytropic equation of state and spinning parallel and antiparallel to the orbital angular momentum. We treat stars with moderate compaction ($(m/R)_{\infty} = 0.14$) and high compaction ($(m/R)_{\infty} = 0.19$). For all but the highest circulation sequences, the spins of the neutron stars increase as the binary separation decreases. Our zero-circulation cases approximate irrotational sequences, for which the spin angular frequencies of the stars increases by $13\%$ ($11\%$) of the
orbital frequency for $(m/R)_{\infty} = 0.14$ ($(m/R)_{\infty} = 0.19$) by the time the innermost circular orbit is reached. In addition to leaving an imprint on the inspiral gravitational waveform, this spin effect is measurable in the electromagnetic signal if one of the stars is a pulsar visible from Earth.
\end{abstract}

\pacs{04.30.Db, 04.25.Dm, 97.80.Fk}

\maketitle


\section{Introduction}
\label{intro}

Binary neutron stars are connected to two fascinating observational phenomena: $\gamma$-ray bursts and gravitational waves. Binary systems of compact objects (neutron stars and black holes) are strong emitters of gravitational radiation and thus prime candidates for the new generation of gravitational wave laser interferometers \cite{Willke:bs,LIGO,Ando:bv,VIRGO,Cutler:2002aa}. Binary neutron stars may also be the engines for $\gamma$-ray bursts of short duration \cite{short_grb}. The numerical modeling of these systems becomes essential for the analysis of observational data in both of these windows.

Numerical simulations of binary neutron stars face two independent challenges: obtaining astrophysically realistic initial data and determining the numerical evolution of such data in time. In this article, we concentrate on the former (initial value) problem and extracting physical insight into the evolution of binaries from these solutions \cite{Baumgarte:2002jm}.

In the past, most researchers \cite{Wilson:1995ty,Wilson:1996ty,bcsst,Mathews:1997pf,Bonazzola:1997gc,Marronetti:1998xv,Bonazzola:1998yq,Marronetti:1998vm,Marronetti:1999ya,Uryu:1999uu,Uryu:2000dw,Usui:2002,Gourgoulhon:2000nn,Taniguchi:2002ns}, have addressed the gravitational part of the initial value problem (IVP) for binary neutron stars (i.e.; the solution of the Hamiltonian and momentum constraints for a quasi-equilibrium circular orbit) via the Wilson-Mathews conformal ``thin-sandwich" approach \cite{Wilson:1995ty, Wilson:1996ty, Cook:2000vr}. This method is typically based on restricting the spatial metric tensor to a conformally flat form and imposing a helical Killing vector to the spacetime to enforce the ``circular orbit" condition \cite{footnote1}. The hydrodynamical part of the IVP however, has been addressed so far in three different ways:

(1) Wilson, Mathews, and Marronetti \cite{Wilson:1995ty, Wilson:1996ty} used an approximate set of relativistic Euler equations with post-Newtonian radiation reaction to relax the fluid motion to steady state. This method is difficult to implement numerically, since it requires a fully-functional, hydrodynamical evolution code.

(2) Baumgarte {\it et al.} \cite{bcsst} solved the Bernoulli equation for a corotating binary. This corresponds to a static fluid in the reference frame rotating with the system, for which the Bernoulli equation applies. While straightforward to implement numerically, corotating systems do not represent astrophysically realistic binaries, due to the high viscosity required to produce the tidal lock \cite{bc92,k92}.

(3) Bonazzola {\it et al.} \cite{Bonazzola:1997gc,Asada:1998,Gourgoulhon:1998dr}, Teukolsky \cite{Teukolsky:1998}, and Shibata \cite{Shibata:1998um} developed a formalism \cite{footnote2} for systems with zero fluid vorticity (irrotational binaries). This formalism requires the solution of an extra elliptic equation for the matter (in addition to the Hamiltonian and momentum constraints). Adopting this method, Bonazzola {\it et al.} \cite{Bonazzola:1998yq}, Marronetti {\it et al.} \cite{Marronetti:1999ya}, Ury\=u and Eriguchi \cite{Uryu:1999uu}, and Taniguchi and Gourgoulhon \cite{Taniguchi:2002ns} obtained numerical solutions for the irrotational binaries. However, the extra elliptic equation requires boundary conditions that have to be specified on the stellar surface, something difficult to implement accurately in Cartesian coordinates. Different investigators have resorted to different techniques to deal with this complication, like pseudo-spectral methods \cite{Bonazzola:1998yq, Taniguchi:2002ns}, decomposition of the elliptic equation into homogeneous and inhomogeneous parts in Cartesian coordinates \cite{Marronetti:1998vm,Marronetti:1999ya}, and overlapping patches of spherical coordinates \cite{Uryu:1999uu,Uryu:2000dw}.

We present in this paper a new hydrodynamical recipe that has as its most important feature the capability of providing solutions to the IVP for binaries with stars with arbitrary spins. This scheme also requires the solution of an extra elliptic equation, but here the method reduces the difficulty attached to implementing stellar surface boundary conditions (particularly complex in Cartesian coordinates). This feature allowed us to obtain accurate solutions using low resolution Cartesian grids.

We construct sequences of quasi-equilibrium orbits with constant rest mass and constant relativistic circulation along the stellar equator. These sequences, which mimic inspiral evolutions outside the innermost stable circular orbit (ISCO), provide insight into the coupling of spin and orbital angular momenta in general relativistic binaries \cite{Price:2001un}.

In Sec. \ref{fields} we discuss our treatment of the gravitational fields, which follows the standard Wilson-Mathews conformal thin-sandwich (CTS) approach. In Sec. \ref{ME} we introduce our new method that allows the free determination of the stellar spins for configurations that satisfy the Euler and continuity equations. The full set of equations is summarized in Sec. \ref{set_equations}. Sections \ref{implementation} and \ref{test} deal with our numerical implementation and code tests, including a comparison of our results with those of Baumgarte {\it et al.} \cite{bcsst} for corotating binaries and Ury\=u and Eriguchi \cite{Uryu:1999uu} for irrotational binaries. Finally, Sec. \ref{results} presents our results for sequences of binaries in quasi-circular orbits for different values of the equatorial circulation, including a comparison with PPN results.


\section{Spacetime Equations}
\label{fields}

\subsection{Initial value problem}
\label{ivp}

In this Sec. we present the equations for the Wilson-Mathews conformal ``thin-sandwich" method \cite{Wilson:1995ty, Wilson:1996ty}, following the notation employed by Baumgarte {\it et al.} \cite{bcsst}. We refer the reader to those papers for a more detailed explanation of the method and the justification of its approximations.

In the (3+1) spacetime formalism of
Arnowitt-Deser-Misner (ADM) \cite{adm62,y79} the line element may be written
\ba 
ds^2 = g_{\mu\nu} dx^\mu dx^\nu = & - & (\alpha^2 - \beta_l\beta^l) dt^2 \nonumber \\
& + & 2 \beta_l dx^l dt + \gamma_{lm}dx^l dx^m~, \nonumber
\ea
where $g_{\mu\nu}$ is the four-metric, $\alpha$ the lapse function, $\beta_i$ the shift vector, and $\gamma_{ij}$ is the spatial three-metric. Latin (Greek) indices go from 1 to 3 (0 to 3), and we set $G=c=1$. The symbols $(~)_{;\mu}$, $D_i$, and $\nabla_i$ will represent the covariant derivatives associated with the tensors $g_{\mu\nu}$, $\gamma_{ij}$, and $\delta_{ij}$ respectively.

The Hamiltonian and momentum constraint equations can be written as
\ba
R - K_{lm}K^{lm} + K^2 &=& 16\pi \rho~, \nonumber \\
D_l(K^{li} - \gamma^{li}K) &=& 8 \pi j^i~,
\label{constraints}
\ea
where $R$ is the three-dimensional Ricci scalar, $K_{ij}$ the extrinsic curvature, and $K$ its trace. The source terms $\rho$ and $j^i$ are the total mass-energy density and the spatial components of the four-momentum density, and they are related to the matter stress-energy tensor $T_{\mu\nu}$ by
\ba
\rho &=& n^\mu n^\nu T_{\mu\nu}~, \nonumber \\
j^i &=& - \gamma^i_{~\nu}n_{\gamma} T^{\nu\gamma}~,
\label{source_terms}
\ea
where $n^{\mu}$ is the vector normal to a spatial hypersurface. Under maximal spatial slicing ($K=0$), which we adopt, the third term and the second term of the left-hand side (LHS) of the Hamiltonian and momentum constraints vanish. We will also define for later use, a source term $S$
\ba
S = \gamma^{lm} T_{lm}~. \nonumber
\ea


\subsection{Reference frames and Killing vectors}
\label{killing_vector}

Binaries in quasi-equilibrium circular orbits are systems that admit an approximate a timelike Killing vector of the form
\ba 
\xi = \partial_t + \Omega ~\partial_\phi~, \nonumber
\ea
in a coordinate frame tied to inertial observers at infinity at rest with respect to the system center of mass (``inertial frame"). Here $\Omega$ is the orbital angular velocity of the binary and $\phi$ the azimuthal coordinate. In Cartesian coordinates, this vector has the components $\xi^\mu = (1,\xi^i)$  with $\xi^i \equiv (\vec{\Omega} \times \vec{r})^i$ ($\vec{r}$ the coordinate radius). No exact Killing vector of this type can exist in astrophysically realistic binaries, since the presence of gravitational radiation leads to an inspiral motion that breaks the symmetry. However the amount of energy dissipated per orbital period (outside the ISCO) is small enough to make this approximation valid in most scenarios \cite{Cook:1995cp,Mathews:1997pf,Miller:2003vw,Duez:2000jt}.

In a rotating frame (i.e., a coordinate system that follows the binary) the Killing vector adopts its simplest form $\xi = \partial_t$. Due to this simplification, we will work in such frame, where the stars are stationary \cite{footnote3}. The relation between the fields in the inertial and rotating frames is given in \cite{Duez:2002bn}.


\subsection{Conformal thin-sandwich equations}
\label{wilson-mathews}

Following \cite{Wilson:1996ty}, the constraint equations (\ref{constraints}) are supplemented by the time evolution equations for the spatial metric
\ba \label{gamma_dot}
\partial_{t} \gamma_{ij} = - 2 \alpha K_{ij} + D_i \beta_{j} + D_j \beta_{i}~,
\ea
where Eq. (\ref{gamma_dot}) can be interpreted as the definition of the extrinsic curvature $K_{ij}$. To simplify the problem the spatial metric $\gamma_{ij}$ is restricted to be conformally flat,
\ba
\gamma_{ij} = \Psi^4  \delta_{ij}~,
\label{conftensor}
\ea
where the conformal factor $\Psi$ is a positive scalar function. For this conformally flat metric, the Ricci scalar $R$ becomes
\ba
R = - 8 \Psi^{-5} \nabla^2 \Psi~, \nonumber
\ea
where $\nabla^2$ is the flat space Laplacian. Using this equation and defining $\tilde K^{ij} \equiv \Psi^{10} K^{ij}$, the Hamiltonian constraint becomes
\ba
\nabla^2 \Psi = - \frac{1}{8} \Psi^{-7} \tilde K_{lm} \tilde K^{lm}    - 2\pi\Psi^5 \rho~.
\label{Psi_Eq}
\ea
To impose the condition of quasi-equilibrium, we set the time evolution of the trace-free part of the spatial metric $\gamma_{ij}$ to be zero. This restriction together with Eq. (\ref{gamma_dot}) and the definition of  $\tilde K^{ij}$ lead to
\ba
\tilde K^{ij} = \frac{\Psi^6}{2\alpha}    \left( \nabla^i \beta^j + \nabla^j \beta^i    - \frac{2}{3} \delta^{ij} \nabla_l \beta^l \right).
\label{K_def}
\ea
Now we can rewrite the momentum constraint equations as a set of elliptic equations for the shift vector
\ba
\nabla^2 \beta^i & + & \frac{1}{3} \nabla^i (\nabla_l \beta^l) = \nonumber \\  & & 2 \nabla_l \ln( \alpha \Psi^{-6} ) \tilde K^{il} + 16 \pi \alpha \Psi^4 j^i~. \nonumber
\ea
These equations are split into two by decomposing $\beta^i$ according to 
\ba \label{beta1}
\beta^i = \frac{1}{4} \nabla^i B - G^i + (\vec{\Omega} \times \vec{r})^i~,
\ea
with the corresponding equations for $B$ and $G^i$
\ba
\nabla^2 B &=& \nabla_l G^l~, \nonumber
\ea
\ba
\nabla^2 G^i &=& -2 \nabla_l \ln( \alpha \Psi^{-6} ) \tilde K^{il} - 16 \pi \alpha \Psi^4 j^i~.
\label{G_Eq}
\ea
Note that the term $(\vec{\Omega} \times \vec{r})^i$ that dominates the asymptotic behavior of the shift vector in the rotating frame has been given explicitly, instead of including it in the fields $G^i$. The reason is simply that this decomposition allows for better control of the boundary conditions for the Eqs. (\ref{G_Eq}) (see Sec. \ref{set_equations} for details).

Finally, an elliptic equation for the product of the conformal factor and the lapse function $(\Psi \alpha)$, is derived using the time evolution equation for the extrinsic curvature $K_{ij}$, by requiring the quasi-equilibrium condition $\partial_{t} K = 0$
\ba
\nabla^2(\alpha \Psi) = \alpha \Psi \left( \frac{7}{8} \Psi^{-8}    \tilde K_{lm} \tilde K^{lm}  + 2 \pi \Psi^4 (\rho + 2 S) \right)~.
\label{alpha_Eq}
\ea


\section{Matter Equations}
\label{ME}

\subsection{Bernoulli equation}
\label{m_structure}

We assume that the matter behaves like a perfect fluid with a stress-energy tensor
\ba
T^{\mu\nu}=\left( \rho_{0}(1+\epsilon)+P \right) \ u^\mu u^\nu + P g^{\mu\nu}~, \nonumber
\ea
with $\rho_{0}$ the proper baryonic mass density, $P$ the pressure, $\epsilon$ the internal energy density, and $u^\mu$ the fluid four-velocity. As described in Sec. \ref{killing_vector}, we will assume that the system is embedded in a four-dimensional manifold endowed with a Killing vector $\xi^\mu$. Following \cite{Problem_Book}, we contract the Euler equations for a perfect fluid with the Killing vector $\xi^\mu$
\ba
(\rho_0 (1+\epsilon)+P) \xi^{\mu} u^{\nu} u_{\mu;\nu} = - \xi^\mu P,_{\mu}
- \xi^\mu u_{\mu} u^{\nu} P,_{\nu}
\label{p1}
\ea

and derive the identity
\ba
\xi^{\mu} u^{\nu} u_{\mu;\nu} & = &u^{\nu} (\xi^\mu u_\mu)_{,\nu} - (u^\nu \xi^\mu_{;\nu}) u_\mu \nonumber \\
& = & u^{\nu} (\xi^\mu u_\mu)_{,\nu}
\label{p2}
\ea
from the antisymmetry of $\xi_{\mu;\nu}$. Knowing that $\xi^\mu P_{,\mu} = 0$, we combine Eqs. (\ref{p1}),(\ref{p2}) 
\ba
\left( \rho_0 (1+\epsilon)+P \right) ~d(u_\mu \xi^\mu)/d\tau = -u_\mu \xi^\mu dP/d\tau~, \nonumber
\ea
which, together with the first law of thermodynamics $d(\rho_0(1+\epsilon)) = \left( \rho_0 (1+\epsilon)+P \right) ~dn_0/n_0$, leads to
\ba
{{d(u_\mu \xi^\mu)} \over {u_\mu \xi^\mu}} = \frac{dn_0} {n_0} -\frac{d \left( \rho_0(1+\epsilon)+P \right) }{\rho_0(1+\epsilon)+P}~, \nonumber
\ea
where $n_0$ is the baryonic number density. Note that $d \left( \ln(u_\mu \xi^\mu) - \ln(n_0) + \ln [\rho_0(1+\epsilon)+P] \right) = 0 $, or in more compact form
\ba \label{bern2}
F =  \frac{\tilde{C} ~n_0}{\rho_0(1+\epsilon)+P} ~,
\ea
where $F \equiv u_\mu \xi^\mu$ and $\tilde{C}$ is a constant. In general $\tilde{C}$ is a constant only along  flow lines and not all across the spatial slice. However, in two special binary systems, namely corotating and irrotational, $\tilde{C}$ has been shown to be a spatial constant 
\cite{Gourgoulhon:2000nn,Shibata:1998um}. (In those two cases, in fact, $\tilde{C}$ is a global spacetime constant since the fluid elements will carry the same value all along the flow lines.) Motivated by the fact that corotating and irrotational binaries describe extreme opposite physical systems, we will take as an approximation $\tilde{C}$ to be constant across a spatial slice for our adopted velocity fields (see Sec. \ref{fluid_motion}) \cite{footnote4}.

Now we will write the field $F$ in a more tractable form. The Killing vector $\xi^\mu$ can always be written as $\xi^\mu = (1,\xi^i)$. Then $F$ becomes
\ba \label{F1}
F = u_\mu \xi^\mu = u_0 + u_l \xi^l~.
\ea
Substituting the contravariant components $u^i$ of the four-velocity
\ba
u_0 & = & g_{0\mu} u^\mu = g_{00} u^0 + g_{0l} u^l~, \nonumber \\
u_i & = & g_{i\mu} u^\mu = g_{i0} u^0 + g_{il} u^l~, \nonumber
\ea
into Eq. (\ref{F1}), and using Eq. (\ref{conftensor}) for the spatial metric, we get
\ba
F = -\alpha^2 u^0 + \Psi^4 \delta_{lm} (\beta^l +\xi^l) (u_0 \beta^m+u^m)~. \nonumber
\ea
A more convenient form of this equation can be obtained by replacing the spatial components of the fluid four-velocity $u^i$ by the components of the fluid coordinate three-velocity $v^i \equiv u^i / u^0$:
\ba \label{F3}
F = u^0 \left[-\alpha^2 + \Psi^4 \delta_{lm} (\beta^l +\xi^l) (\beta^m+v^m) \right]~.
\ea
The normalization condition for the four-velocity
\ba \label{norm_vel}
u_\mu u^\mu=-1~,
\ea
leads to the following expression for $u^0$
\ba \label{u0}
u^0 = \left[ \alpha^2 - \Psi^4 \delta_{lm} (\beta^l + v^l) (\beta^m + v^m) \right]^{-1/2}~.
\ea
Inserting this expression into Eq. (\ref{F3}) we finally arrive at the formula
\ba \label{F4}
F = - \frac{\alpha^2 - \Psi^4 \delta_{lm} (\beta^l +\xi^l) (\beta^m+v^m)}
{\left[\alpha^2 - \Psi^4 \delta_{lm} (\beta^l + v^l) (\beta^m + v^m)\right]^{1/2}}~,
\ea
where $F$, and thus Eq. (\ref{bern2}), is a function of the gravitational and gauge fields $\Psi$, $\alpha$, and $\beta^i$, Killing vector spatial components $\xi^i$, and the coordinate three-velocity of the fluid $v^i$.


\subsection{Continuity equation}
\label{cont_eq}

In the previous Sec. we obtained the Bernoulli equation (\ref{bern2}) assuming the existence of a timelike Killing vector and using the Euler equation for a perfect fluid. Now, we need to guarantee that the fluid will also satisfy baryonic mass conservation
\ba
(\rho_0 u^\mu)_{;\mu} = 0~.
\label{cont1}
\ea
In the rotating frame, Eq. (\ref{cont1}) can be explicitly written for a spatially conformally flat spacetime as (see Appendix \ref{appendix_Cont} for the derivation)
\ba \label{cont3}
\nabla_l(\rho_0 v^l_R) + \rho_0 v^l_R \nabla_l \left( \ln (u^0 \alpha \Psi^6) \right) = 0~,
\ea
where
\ba \label{vel_rel}
v^i_R = v^i_I - (\vec{\Omega} \times \vec{r})^i~,
\ea
and where R (I) labels a quantity in the rotating (inertial) reference frame.
We can always decompose the three-velocity $v^i_R$ as the sum of a solenoidal and an irrotational part
\ba
v^i_R = v^i_{RS} + v^i_{RI}~,
\label{decomp1}
\ea
where
\ba
\nabla_l  v^l_{RS} = 0~,
\label{decomp2}
\ea
and
\ba
\epsilon_{ilm} ~ \nabla^l v^m_{RI} = 0~,
\label{decomp3}
\ea
and where $\epsilon_{ilm}$ is the three-dimensional Levi-Civita tensor.
The solenoidal component $v^i_{RS}$ will be specified to force the bulk of the fluid to move in a pre-determined way (see Sec. \ref{fluid_motion} for more details). The irrotational component $v^i_{RI}$ on the other hand, will be determined by satisfying the continuity equation (\ref{cont3}). In order to derive an equation for $v^i_{RI}$, we will define a scalar function $\sigma$ such that
\ba
v^i_{RI} = \nabla^i \sigma~.
\label{sigma}
\ea
The existence of this scalar is guaranteed by the condition (\ref{decomp3}). Inserting Eq. (\ref{decomp1}) into the continuity equation (\ref{cont3}), we get
\ba
\rho_0 ~\nabla_l (v^l_{RI}) & &+ ~ (v^l_{RS}+v^l_{RI}) \nonumber\\
& & \left( \nabla_l \rho_0 + \rho_0 ~\nabla_l [ \ln(u^0 \alpha \Psi^6)] \right)
= 0~. \nonumber
\ea
We divide now by $\rho_0$ (which is non-zero inside the star) and collect the terms in the second parenthesis of the second term
\ba
\nabla_l (v^l_{RI}) &+& (v^l_{RS}+v^l_{RI}) ~\nabla_l \left( \ln(\rho_0 u^0 \alpha \Psi^6) \right) = 0~. \nonumber
\ea
Using the definition of Eq. (\ref{sigma}), we get an elliptic equation for $\sigma$
\ba
\nabla^2 \sigma = -(v^l_{RS}+\nabla^l \sigma) ~\nabla_l \left( \ln(\rho_0 u^0 \alpha \Psi^6) \right)~.
\label{sigma_eq}
\ea
This equation has to be solved in conjunction with an appropriate boundary condition. A typical choice is a condition that forces the fluid velocity at the stellar surface to be tangent to this surface, like the ones employed in Newtonian \cite{Bonazzola:1992} and relativistic \cite{Teukolsky:1998} irrotational binaries. This condition can be written, for instance, as
\ba
v^l_R ~\nabla_l\rho_0 |_{Surf} = (v^l_{RS}+\nabla^l \sigma) ~\nabla_l \rho_0 |_{Surf} = 0~.
\label{sigma_bc}
\ea
A condition of this type becomes problematic in Cartesian coordinates, since it involves imposing a Neumann-like condition on a near-spherical surface. Marronetti {\it et al.} \cite{Marronetti:1999ya} solved the problem by splitting the elliptic equation for the potential field into homogeneous and inhomogeneous parts, and letting the homogeneous field take care of the boundary condition. While formally correct, this solution is very difficult to implement accurately and the results are strongly dependent on grid resolutions \cite{Marronetti:1999ya}. Because of boundary conditions like this, other groups resorted to spectral methods \cite{Bonazzola:1997gc,Gourgoulhon:2000nn,Taniguchi:2002ns} or superposing mappings of spherical polar coordinates \cite{Uryu:1999uu,Uryu:2000dw}.

This problem can be minimized in the following way. Let us introduce the field $\lambda \equiv \sigma \rho_0$. Knowing that
\ba
\nabla_l(\rho_0 \nabla^l \sigma) = \nabla^2 (\sigma \rho_0) - \nabla_l(\sigma \nabla^l\rho_0)~, \nonumber
\ea
we can write the first term of Eq. (\ref{cont3}) as
\ba
\nabla_l(\rho_0 v^l_R) & = & \nabla_l(\rho_0 v^l_{RS}) + \nabla_l(\rho_0 \nabla^l \sigma) \nonumber \\
& = & \rho_0 \nabla_l v^l_{RS} + v^l_{RS} \nabla_l\rho_0  \nonumber \\
& + & \nabla^2 (\sigma \rho_0) - \nabla_l(\sigma \nabla^l\rho_0)~. \nonumber
\ea
Going back to Eq. (\ref{cont3}), using Eq. (\ref{decomp2}), and rearranging the terms, we get an elliptic equation for the field $\lambda$
\ba
\nabla^2 \lambda & = & \nabla_l(\sigma ~\nabla^l\rho_0) - \rho_0 v^l_R ~ \nabla_l \left( \ln(u^0 \alpha \Psi^6) \right) \nonumber \\
& - & v^l_{RS} ~\nabla_l \rho_0~, \nonumber
\ea
which, when writing the RHS in terms of $\lambda$, becomes
\ba
\nabla^2 \lambda & = & \nabla_l(\lambda ~\nabla^l \ln(\rho_0)) - v^l_{RS} ~\nabla_l \rho_0 + \nabla_l \ln(u^0 \alpha \Psi^6) \nonumber \\
& & (\nabla^l \lambda - \rho_0 v^l_{RS} - \lambda~\nabla_l  \ln(\rho_0))~. \nonumber
\ea
To determine a boundary condition for this equation, we write the condition for $\sigma$ (\ref{sigma_bc}) as a function of the new field $\lambda$
\ba
\nabla^l (\frac{\lambda}{\rho_0}) ~\nabla_l \rho_0 |_{Surf} = -v^l_{RS} ~\nabla_l \rho_0 |_{Surf}~. \nonumber
\ea
The RHS of this equation is zero if the stellar surface is spherical and the field  $v^i_{RS}$ is specified by Eq. (\ref{vrs}) given below. This will not be the case in reality, since tidal forces and spin will deform the surface away from sphericity. However, the spherical approximation was employed in previous work \cite{Marronetti:1998vm}, where it was shown to have little impact on the final results. Approximating the RHS of the above equation to zero, we can satisfy the resulting boundary condition by imposing 
\ba \label{lambda_bc}
\lambda|_{Surf} = \nabla^l \lambda|_{Surf}  = 0
\ea 
in a vicinity of the stellar surface. For instance, this condition can be numerically imposed on the grid points where $\rho_0$ falls below some threshold value (see Sec. \ref{set_equations}). This approximate boundary condition is easier to implement in Cartesian coordinates than the Neumann condition (\ref{sigma_bc}).


\subsection{Bulk fluid motion}
\label{fluid_motion}

The bulk of the stellar fluid motion in the rotating frame resides in
the field $v^i_{RS}$. Since we are free to specify $v^i_{RS}$ in any way that satisfies Eq. (\ref{decomp2}), we will choose the form
\ba \label{vrs}
v^i_{RS} = (a-1) \left[ \vec{\Omega} \times (\vec{r}-\vec{r}_0) \right]^i~,
\ea
where $a$ is a free parameter, $\vec{\Omega}$ is the angular three-velocity of the binary system, $\vec{r}$ is the coordinate position vector, and $\vec{r}_0$ is the geometrical center of the star (determined by the point of maximum baryonic density). This condition forces the bulk of the fluid to move like a rigid rotator with angular velocity parallel to the orbital angular velocity $\vec{\Omega}$ and magnitude $(a-1) ~\Omega$ in the rotating frame. The fluid three-velocity in the inertial frame $v^i_I$ is related to its counterpart in the rotating frame $v^i_R$ by Eq. (\ref{vel_rel}) which, together with Eqs. (\ref{decomp1}), (\ref{sigma}), and (\ref{vrs}) gives
\ba
v^i_I = (\vec{\Omega} \times \vec{r}_0)^i + a~[\vec{\Omega} \times (\vec{r}-\vec{r}_0)]^i + \nabla^i \sigma~.
\label{vI}
\ea
A star with this velocity profile will spin at a frequency (as seen by a distant observer in the inertial frame) $\vec{\Omega}_s = a ~\vec{\Omega}$. Thus, we will refer to $a$ as the {\it spin parameter} of the stars. If $a=1$ then $v^i_{RS}$ is zero, and $\sigma= const$ is the exact solution to Eq. (\ref{sigma_eq}), with boundary conditions (\ref{sigma_bc}). Then $v^i_I$ becomes $(\vec{\Omega} \times \vec{r})^i$, which is the condition that defines corotating binary systems. In appendix \ref{appendix_Bern} we show that for this particular case, the Bernoulli equation (\ref{bern3}) assumes the standard form for corotating binaries [see for instance Eq. (32) of Baumgarte {\it et al.} \cite{bcsst}]. Thus, setting $a=1$ we recover the full set of CTS equations corresponding to corotating binaries.

The spin parameter $a$ allows us to control the stellar spin, while the scalar field $\sigma$ provides the correction necessary to satisfy baryonic mass conservation. Perfect fluid irrotational binaries have been obtained in the past by requiring the fluid vorticity tensor to be zero \cite{Teukolsky:1998,Shibata:1998um}
\ba
\omega_{\mu\nu} \equiv (h u_\nu)_{;\mu} - (h u_\mu)_{;\nu}  = 0~. \nonumber
\ea
For these cases, the flow $hu^\mu$ can be written as the gradient of a potential field, for which a new elliptic equation is derived. The method we propose here does not enforce this condition on the fluid. Our method will rely on the concept of circulation to approximate solutions with zero vorticity. The Kelvin-Helmoltz theorem \cite{c79} states that the relativistic circulation
\ba \label{circ1}
{\cal C}(c) = \oint_{c} h u_{\mu} d\sigma^{\mu}
\ea
is conserved for isentropic fluids following any closed path $c$, when evaluated on hypersurfaces of constant proper time. Integral conservation laws like the Kelvin-Helmholtz theorem can be defined more generally than on hypersurfaces of constant proper time. For instance, in a reference frame where exists a timelike  Killing vector whose coordinates are 
$\xi^\mu =(1,\vec{0})$ , it can be shown than $\partial{\cal C}(c) / \partial t = 0$  \cite{c79}. This simply states the fact that in a frame where the fluid is stationary (like the frame corotating with the binary) the circulation is also time independent. Keeping the circulation constant from one orbit in the sequence to another is our working hypothesis. This is consistent with other assumptions of the quasi-equilibrium method, like assuming that the radial velocity of the stars remains zero all along the orbits of the sequence. In reality, the radial velocity like the circulation on t=const hypersurfaces will change slightly when the stars get closer to each other and full time numerical evolutions will eventually estimate qualitatively this effect. In this paper we will evaluate the integral (\ref{circ1}) along the stellar equator and this is what will be referred to as the circulation of the star. The two stars in the inspiraling binary will conserve the value of circulation ${\cal C}_\infty$ they had at large separation. The circulation of a single irrotational star (i.e., at rest in the inertial frame) is zero. Consequently, we will represent an irrotational sequence by the quasi-stationary CTS solutions for stars with ${\cal C} = 0$. Note that conservation of circulation will also be true for sequences with ${\cal C} = {\cal C}_\infty \neq 0$.

In the Newtonian limit, the value of $a=0$ will correspond to ${\cal C} = 0$ and in Appendix \ref{appendix_Newt} we show that in this case our scheme recovers the exact equations corresponding to Newtonian irrotational binaries. In Sec. \ref{results} we will show that, for the general relativistic case, our sequences with ${\cal C} = 0$ agree very well with the irrotational binary sequences obtained using an exact treatment. Of course, by adopting the form of the velocity field according to Eq. (\ref{vI}) we are making an approximation to the exact irrotational velocity field. In the Newtonian limit, our velocity field requires $a=0$ to have zero vorticity.

Finally, we note that Eq. (\ref{vI}) does not represent the only type of fluid motion that our method would allow. We could also specify a differential rotation velocity profile, by modifying the scheme presented in this paper to accommodate such circumstances.


\subsection{Equation of state}
\label{EOS}

The method described in the previous sections is independent of the equation of state (EOS) used to model the stellar matter. In this paper and for the sake of simplicity, we will employ a polytropic EOS
\ba \label{eos}
P = \kappa ~\rho_{0}^{\Gamma}~,
\ea
where $\kappa$ is the polytropic constant and $\Gamma$ the adiabatic index related to the polytropic index $n$ by $\Gamma = 1 + 1/n$. All the results presented here are for $\Gamma=2$, which facilitates a comparison with previous work (see Sec. \ref{results}). 

For polytropic EOS like Eq. (\ref{eos}), the specific enthalpy
\ba
h = \exp \left( \int \frac{dP}{\rho_0 (1 +\epsilon) +P} \right)~, \nonumber
\ea
becomes
\ba
h=1+ \epsilon +\frac{P}{\rho_0}~. \nonumber
\ea
Note that this expression, combined with Eq. (\ref{bern2}) provides a very compact form for the Bernoulli equation
\ba \label{bern3}
F = \frac{C}{h}~,
\ea
where $C$ is a new constant \cite{bcsst}. 

It will be convenient to define a variable $q \equiv \rho_0 /P$ and write every quantity as function of $q$ \cite{bcsst}
\ba
\rho_0 &=& \kappa^{-n} q^n~, \nonumber \\
h &=& q ~(n+1) + 1~, \nonumber \\
\epsilon &=& n~q~, \nonumber \\
P &=& \kappa^{-n} q^{n+1}~. \nonumber
\ea
The matter source terms can then be written as
\ba
\rho &=& \rho_0 ~(1+q(1+n)) ~(\alpha u^0)^2 - P ~, \nonumber \\
j^i &=& \rho_0 ~(1+q(1+n)) ~{u^0}^2 \alpha ~(v^i_R+\beta^i) ~, \nonumber \\
S &=& 3 P + \rho_0 ~(1+q(1+n)) ~[(\alpha u^0)^2 - 1]~. \nonumber
\ea 

Using these relations and the normalization condition for the fluid four-velocity (\ref{norm_vel}) we can derive the expressions needed for the RHS of the elliptic equations (\ref{Psi_Eq}), (\ref{G_Eq}), and (\ref{alpha_Eq})
\ba
\rho &=& \kappa^{-n} q^n \left[ (1+q(1+n))~(\alpha u^0)^2 - q \right]~, \nonumber \\
j^i &=& \kappa^{-n} q^n (1+q(1+n))~\alpha~{u^0}^2 (v_R^i+\beta^i)~, \nonumber \\
\rho + 2S &=& \kappa^{-n} q^n \nonumber \\ & & \left[ (1+q(1+n))~(3~(\alpha u^0)^2 - 2) + 5 q \right]~. \nonumber
\ea


\section{Final Set of Equations}
\label{set_equations}

\begin{table}
\caption{Boundary conditions for the outer boundaries ($r \rightarrow \infty$) and on the coordinate planes in Cartesian coordinates. The stars are aligned along the $y$ axis with a coordinate separation of $d$ and the equatorial plane is $z = 0$.}
\begin{center}
\begin{tabular}{lccc}
\hline
\hline \\
$r \rightarrow \infty$~~~~~~~~~~&~~~~$x = 0$~~~~&~~~~$y = 0$~~~~&~~~~$z = 0$~~~~\\
\tableline
$\Psi - 1 \sim \displaystyle f(r)$ &
$even$  & $even$    &$even$\\[3mm]
$(\alpha \Psi)- 1 \sim \displaystyle f(r)$ &
$even$  & $even$    &$even$\\[3mm]
$G^x \sim \displaystyle \frac{y}{r^3}$ & 
$even$  & $odd$     &  $even$ \\[3mm]
$G^y \sim  \displaystyle \frac{x}{r^3}$ &
$odd$   & $even$    & $even$ \\[3mm]
$G^z \sim \displaystyle \frac{xyz}{r^7}$ &
$odd$   & $odd$ & $odd$\\[3mm]
$B \sim \displaystyle \frac{xy}{r^3}$ &
$odd$   & $odd$ & $even$\\[3mm]
$f(r) \equiv \frac{1}{r}+\frac{d^2}{8~r^5} (3y^2-r^2)$ \\[3mm]
\hline
\hline 
\end{tabular}
\end{center}
\label{Table_1}
\end{table}

In order to simplify the numerical problem, we find useful to nondimensionalize all the equations by factoring out the polytropic constant $\kappa$. Following \cite{bcsst}, we use dimensionless coordinates $\bar{x}^i = \kappa^{-n/2} x^i$, derivative operators $\bar{\nabla}^i = \kappa^{n/2} \nabla^i$, masses $\bar{M} = \kappa^{-n/2} M$, angular velocities $\bar{\Omega} = \kappa^{n/2} \Omega$, angular momenta $\bar{J} = \kappa^{-n} J$, etc. This is equivalent to assuming a value of $\kappa=1$ and omitting the bars on top of the new variables and operators. In the remainder of the paper, all the variables and quantities will be given in these dimensionless units. Results for arbitrary values of $\kappa$ can be determined by applying the above scaling relations.

In the previous sections we derived a set of CTS equations that define our binary system. The fields that will play the role of independent variables are the conformal factor $\Psi$, the product of the conformal factor and the lapse function $(\alpha \Psi)$, the vector field $G^i$ and the scalar $B$, from which the shift vector $\beta^i$ is derived, the flow potential $\lambda$ and the matter density variable $q$. The corresponding equations are
\ba \label{final_eqs}
\nabla^2 \Psi & = & - \frac{1}{8} \Psi^{-7} \tilde K_{lm} \tilde K^{lm} - 2 \pi\Psi^5 q^n \nonumber \\
& & \left[ (1+q(1+n))~(\alpha u^0)^2 - q \right]~,\nonumber \\
\nabla^2 (\alpha \Psi) & = &\frac{7}{8} \alpha \Psi^{-7} \tilde K_{lm} \tilde K^{lm} + 2 \pi \alpha \Psi^5  q^n  \nonumber \\
& & \left[ (1+q(1+n))~(3 (\alpha u^0)^2 - 2) + 5 ~q \right]~,\nonumber \\
\nabla^2 G^i & = & -2 \nabla_l \ln( \alpha \Psi^{-6} ) \tilde K^{il} - 16 \pi \Psi^4 q^n \nonumber \\
& & (v_R^i + \beta^i)~(1+q(1+n))~(\alpha u^0)^2 ~, \nonumber \\
\nabla^2 B &=& \nabla_l G^l~,\nonumber \\
\nabla^2 \lambda &=& n \nabla_l(\lambda ~\nabla^l \ln(q)) -v^l_{RS} ~\nabla_l q^n \nonumber \\
& + & (\nabla^l \lambda - q^n v^l_{RS} - \lambda~n~\nabla^l  \ln(q)) \nonumber \\ & & \nabla_l \ln(u^0 \alpha \Psi^6)~,
\ea 
and the Bernoulli equation
\ba \label{bern4}
q = { 1 \over n+1} \left({C \over F} - 1 \right)~.
\ea
These equations are completed with the definition of $\tilde K^{ij}$ (\ref{K_def}), the formulas that define the relation between the shift vector $\beta^i$, $G^i$, and $B$ (\ref{beta1}), the field $F$ (\ref{F4}), and the velocity field $v^i_{RS}$ (\ref{vrs}).

We work with binary systems composed of identical stars aligned along the $y$ axis with $z=0$ as the orbital plane. The symmetries of these systems allow us to solve the equations in one octant of the numerical grid. Equations (\ref{final_eqs}) form a set of seven nonlinear, coupled elliptic equations that is solved numerically as described in Section \ref{implementation}. The boundary conditions for the first six equations are specified at large distances (i.e., at the outer boundary of the numerical grid). They follow a Robin-like algorithm based on the fall-off behavior of the lowest order non-vanishing multipole moments for the fields $G^i$ and $B$, and the first and second non-vanishing multipole moments for the fields $\Psi$ and $(\alpha \Psi)$. The quadrupole moments for $\Psi$ and $(\alpha \Psi)$ are required in order to get an accurate measure of the gravitational (ADM) mass, due to the close proximity of the boundaries. The details of the fall-off dependence for each field, as well as the reflection symmetries at each plane, are given in Table \ref{Table_1}. The elliptic equation for the field $\lambda$ requires $\lambda=\nabla^i \lambda=0$ at the surface of the star. This condition is imposed by forcing $\lambda=0$ whenever the baryonic density falls below some threshold value, which is usually set at $5\%$ of the maximum stellar density, to avoid problems related to the steep gradients of the density right at the stellar surface. This threshold can be dropped to smaller values by increasing the grid resolution.

In Appendix \ref{appendix_int} we derive the formulas used to calculate the rest mass $M_0$, the ADM mass $M$, the angular momentum $J$, and the circulation $\cal{C}$ of each quasi-stationary orbit.  


\section{Numerical Implementation}
\label{implementation}

\subsection{Elliptic solver}
\label{ell_solver}
The core of the numerical code used to solve the set of equations ({\ref{final_eqs}) is an elliptic solver. This solver is based on a variation of a finite difference multigrid algorithm like the one used for previous work \cite{Marronetti:1998xv,Marronetti:1999ya}. The solver makes use of the octant symmetry of the problem and works in the Sec. of the physical space with positive values for the coordinates $x$, $y$, and $z$. Reflection symmetries and boundaries conditions are described in Sec. \ref{set_equations}. For the results presented in this article, three different grid sizes were employed: small ($32^3$), medium ($64^3$), and large ($128^3$ grid points). Larger grid sizes might be needed for the generation of initial data sets for evolutionary codes and will be discussed in future work \cite{Marronetti:2003aa}. A more detailed description of the iterative algorithm at the core of the solver is provided in Appendix \ref{appendix_solver}. Computing the solution of Eqs. ({\ref{final_eqs}) for a single separation distance on the large grid takes typically around 50 CPU hours on the IBM Regatta p690 located at the National Computational Science Alliance (NCSA).


\section{Code Tests}
\label{test}

\begin{figure}
\epsfxsize=2.8in
\begin{center}
\leavevmode \epsffile{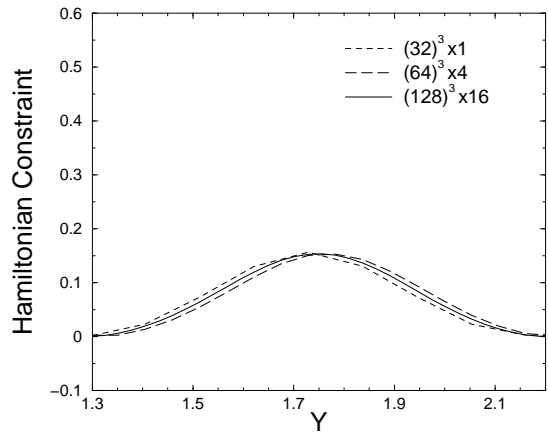}
\end{center}
\caption{ Hamiltonian constraint residual across one of the stars. The residual is evaluated along the line (0.22, y, 0.22). This line is chosen off-centered to avoid the cancellation of some of the constraints due to the symmetries of the problem. The residuals have been plotted for three different grid resolutions: $32^3$ (dotted line), $64^3$ (dashed line), and $128^3$ (solid line). }
\label{HC_conv}
\end{figure}

\begin{figure}
\epsfxsize=2.8in
\begin{center}
\leavevmode \epsffile{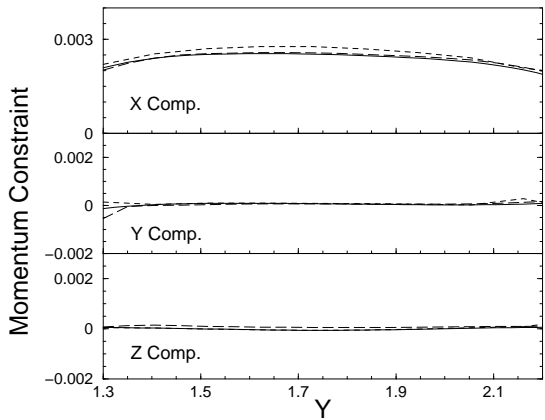}
\end{center}
\caption{ Momentum constraint residuals across one of the stars. The residuals are evaluated along the same line as in Fig. \ref{HC_conv}. Line captions are identical to those of Fig. \ref{HC_conv}. }
\label{MC_conv}
\end{figure}

\begin{figure}
\epsfxsize=2.8in
\begin{center}
\leavevmode \epsffile{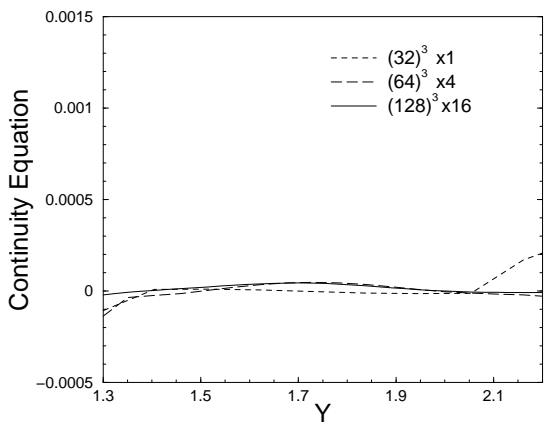}
\end{center}
\caption{ Continuity equation residual across one of the stars. The residuals are evaluated along the same line as in Fig. \ref{HC_conv}. Line captions are identical to those of Fig. \ref{HC_conv}.}
\label{Cont_conv}
\end{figure}

We tested the code solving the set of equations (\ref{final_eqs}) and the Bernoulli equation (\ref{bern4}) for a binary with identical stars with individual rest mass $m_0 = 0.1469$, total mass-energy $m = 0.1368$, and compaction ratio $(m/R)_{\infty} = 0.14$ \cite{footnote5}. The stars were set along the $y$ axis, at a coordinate separation between centers of $d = 3.50$, where the stellar center is defined as the point of maximum baryonic mass density $\rho_0$. The system rotates counterclockwise in the orbital plane $z=0$ and the spin parameter $a$ is set to be zero. This test problem was solved for three different grid sizes with $32^3$, $64^3$, and $128^3$ grid points respectively. These grid sizes correspond to resolutions of about 15, 30, and 60 grid zones across the stellar diameter. In order to check the second order convergence of the code, we plotted the residuals of the Hamiltonian constraint (Fig. \ref{HC_conv}), the components of the momentum constraint (Fig. \ref{MC_conv}), and the continuity equation (Fig. \ref{Cont_conv}), all evaluated in the stellar interior. The residuals are defined as the absolute value of the RHS minus the LHS of Eqs. (\ref{constraints}) and (\ref{cont1}) respectively. The plots show their values along an off-centered line defined by the coordinates $(0.22,~y,~0.22)$ that crosses the upper star from side to side. This off-centering was done to prevent cancellations in the constraints due to the symmetry of the system. 

Figure \ref{Cont_vR} shows the contour plot of the baryonic density and the fluid three-velocity $v^i_R$ in the rotating frame, while Fig. \ref{Cont_vRI} shows the irrotational component of the velocity $v^i_{RI}$ (here $M_0=2 m_0$). As we anticipated in Sec. \ref{fluid_motion}, most of the fluid motion follows the solenoidal field $v^i_{RS}$, determined by Eq. (\ref{vrs}). By setting $v^i_{RS}$, we fixed the dominant part of the fluid velocity, while letting the field $v^i_{RI}$ play the role of small corrections that guarantee satisfaction of the continuity equation (\ref{cont1}). Note that Fig. \ref{Cont_vRI} also highlights another convenient feature of our method: the fact that the field $\lambda$, which is the most difficult numerically, contributes minimally to the final results for the cases explored here. While the velocity field $v^i_{RI}$ is at its largest near the surface, the matter current $\rho_0~v^i_{RI}$ is minimal due to the very steep decay of the rest mass density with increasing radial distance (see contour lines in Fig. \ref{Cont_vRI}). To verify that only small fractions of the total mass and angular momentum of the system are associated with the field $v^i_{RI}$, we compare the main parameters of two identical orbits, one of which was solved neglecting completely the flow potential $\lambda$ (i.e., $\lambda = 0$ everywhere). The difference between the correct numerical results and the ones without $\lambda$ is of the order of a few percent at most. This result justifies our approximate boundary condition for the scalar potential $\lambda$ (\ref{lambda_bc}) (see Table \ref{Table_comp}).

\begin{table}
\begin{center}
\caption{Comparison between the correct numerical solution for a binary with stellar rest mass $m_0 = 0.1469$, coordinate separation $d=4.12$, and circulation ${\cal C}=0$, and the approximate solution obtained when the flow potential $\lambda$ is neglected. As expected, the influence of the field $\lambda$ in the final results is small.}
\label{Table_comp}
\begin{tabular}{lccc} \\
\hline
\hline \\
Parameter~~&~~Correct Result~~&~~$\lambda = 0$ Result~~&~~Diff. (\%)~~\\[2mm]
\tableline\\[0.5mm]
$\Omega ~m_0$ & 0.00858 & 0.00843 & ~1.7 \\[2mm]
$M$ & 0.27174 & 0.27160 & $<$ 0.1 \\[2mm]
$J$ & 0.08494 & 0.08268 & ~2.7 \\[2mm]
Max. $\rho_{0}$ & 0.12375 & 0.12381 & $<$ 0.1 \\[2mm]
$a$ & 0.58 & 0.60 & ~3.3 \\
\hline
\hline
\end{tabular}
\end{center}
\end{table}

Numerically, our solutions obtained in Cartesian coordinates are as accurate as the more sophisticated (but more complicated) alternative numerical methods that employ patches of spherical coordinates \cite{Uryu:1999uu}. These solutions still do not reach the very high accuracy of the pseudo-spectral methods \cite{Bonazzola:1997gc,Taniguchi:2002ns}, but Cartesian coordinates methods usually behave better in the presence of cusps at the stellar surfaces (when the stars are very close to each other). Cartesian coordinates are also more convenient when providing initial data for evolutionary codes that work on those coordinates, since interpolating the fields from one coordinate system to another usually introduces noise into the data set \cite{Marronetti:2003aa}. Figure \ref{Cont_rho_cusp} shows the rest-mass contour plot for the closest orbit of the sequence $(m/R)_\infty=0.14$ and ${\cal C}=0$ \cite{footnote6}. No equilibrium solutions were found inside this orbit.

\begin{figure}
\epsfxsize=2.8in
\begin{center}
\leavevmode \epsffile{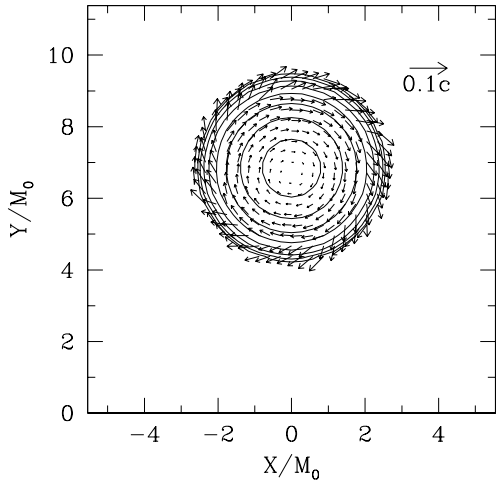}
\end{center}
\caption{ Contour plot of the full fluid (coordinate) three-velocity $v^i_R$ and the rest-mass density $\rho_0$ in the rotating frame. The contour lines are drawn for 
$\rho_0=10^{-(0.2~j+0.1)} \rho_0^{Max} $, where $\rho_0^{Max}$ denotes the maximum value of the rest-mass density $\rho_0$ (here it is $0.1241$), 
for $j=0,1...,7$. Vectors indicate the local velocity field and the scale
is as shown in the top right corner. The coordinates have been normalized by the total rest mass ($M_0 = 0.2938$).}
\label{Cont_vR}
\end{figure}

Friedman {\it et al.} \cite{Friedman:2001pf} derived a virial relation for quasi-equilibrium spacetimes like the ones described here, showing that the virial theorem is satisfied if the gravitational mass of the system is identical to the Komar mass. In the solutions presented in the next section, the relative difference between these masses is of about $2\%$. This difference is due to the proximity of the grid outer boundaries, since both masses are only defined asymptotically. Test runs performed with various grid sizes show convergence of the relative difference to zero. We also find that the quasi-equilibrium relation $\Omega = dM/dJ$ along the equilibrium sequences is satisfied to within $8\%$.

\begin{figure}
\epsfxsize=2.8in
\begin{center}
\leavevmode \epsffile{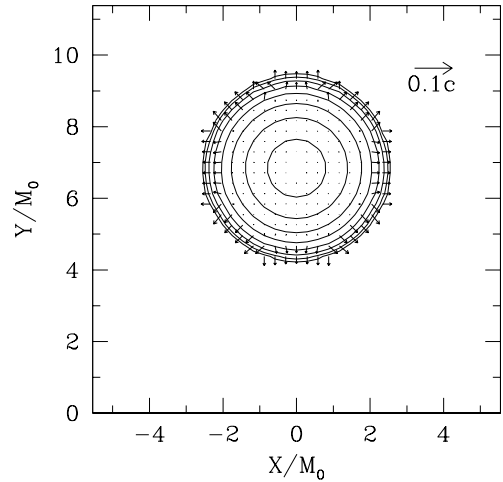}
\end{center}
\caption{ Contour plot of the irrotational component $v^i_{RI}$ of the fluid coordinate three-velocity and the rest-mass density in the rotating reference frame. The plot is labeled as in Fig. \ref{Cont_vR}.}
\label{Cont_vRI}
\end{figure}

\begin{figure}
\epsfxsize=2.8in
\begin{center}
\leavevmode \epsffile{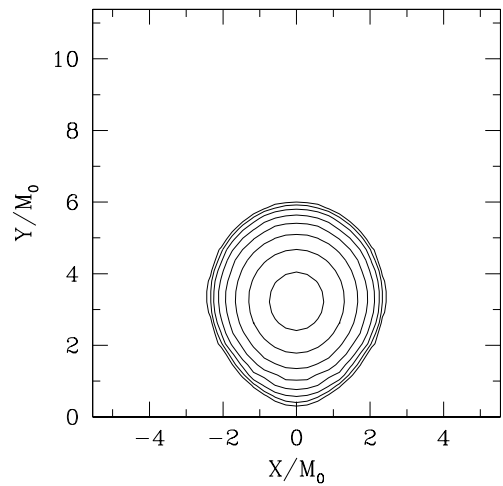}
\end{center}
\caption{ Contour plot of the rest-mass density $\rho_0$ for the closest orbit of the $(m/R)_\infty=0.14$ and ${\cal C}=0$ sequence. The plot is labeled as in Fig. \ref{Cont_vR}.}
\label{Cont_rho_cusp}
\end{figure}


\section{Results}
\label{results}

All the binary sequences described in this paper where modeled using a polytropic EOS with index n=1 ($\Gamma = 2$). For this particular EOS, the critical rest mass (gravitational mass) of a star in isolation is $m_0 = 0.180$ ($m = 0.164$) with a compaction ratio of $(m/R)_\infty = 0.216$ \cite{footnote7}. All the binaries are composed of identical stars and the orbital plane is $z=0$. 

We study models that have two different compaction ratios in isolation: a moderate value $(m/R)_{\infty} = 0.14$ and a high value $(m/R)_{\infty} = 0.19$. These compaction ratios correspond to individual stars with rest masses $m_0=0.1469$ and $m_0=0.1767$, respectively. When these stars are spun up to the mass-shedding frequency, their equatorial circulation assumes the critical values ${\cal C}_{crit}=1.96$ and ${\cal C}_{crit}=2.02$, respectively, values which will be used to normalize the circulations of the binary sequences in the next section.


\subsection{Comparison with previous work}

\begin{figure}
\epsfxsize=2.8in
\begin{center}
\leavevmode \epsffile{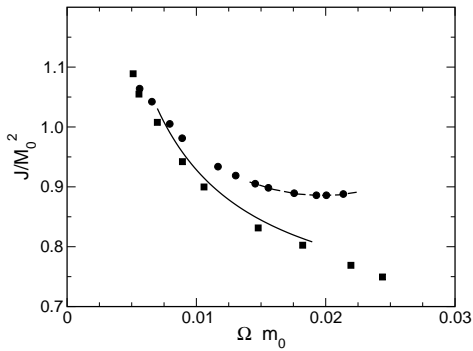}
\end{center}
\caption{ Angular momentum vs orbital frequency. We compare a sequence of corotating orbits for stars with compaction ratio at infinity $(m/R)_\infty=0.15$ from Baumgarte {\it et al.} \cite{bcsst} (dashed line) with our sequence with $a = 1.0$ (filled circles). For the irrotational case, we compare a sequence of orbits for stars with compaction ratio at infinity $(m/R)_\infty=0.14$ from Ury\=u and Eriguchi \cite{Uryu:1999uu} (solid line) with our sequence with null circulation (filled squares).}
\vspace {4 mm}
\label{J_Comparison}
\end{figure}

\begin{figure}
\epsfxsize=2.8in
\begin{center}
\leavevmode \epsffile{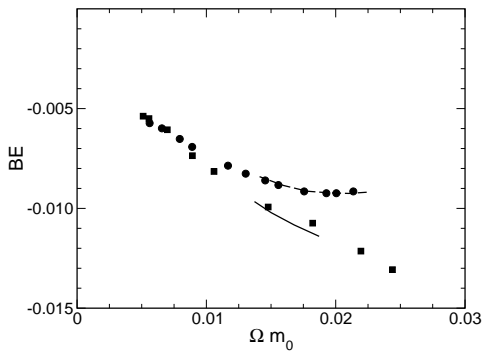}
\end{center}
\caption{ Binding energy comparison between our results and previous work. The binding energy is defined as $BE \equiv (M - M_\infty)/M_0$, where $M$ is the gravitational (ADM) mass. The notation is identical to that of Fig. \ref{J_Comparison}.}
\label{BE_Comparison}
\end{figure}

We tested our code by reproducing a corotating sequence ($a=1$) identical to one provided by Baumgarte {\it et al.} \cite{bcsst} for stars with compaction ratio $(m/R)_\infty = 0.15$. We also compared one of our ${\cal C}=0$ sequences with the irrotational counterpart given by Ury\=u and Eriguchi \cite{Uryu:1999uu} for stars with compaction ratio $(m/R)_\infty = 0.14$. Figures \ref{J_Comparison} and \ref{BE_Comparison} show the total angular momentum and binding energy versus the orbital angular velocity. The agreement between our $a=1$ sequence and the corotating sequence from \cite{bcsst} further validates our numerical code. Most significantly is the close agreement between our ${\cal C} = 0$ sequence and the irrotational sequence from \cite{Uryu:1999uu}, given that the two sequences were obtained using different (albeit closely related) formalisms and numerical approaches. The overlapping system of spherical coordinates used in  \cite{Uryu:1999uu} is better suited than the Cartesian grids used here for dealing with the boundary conditions of the elliptic equation for the fluid velocity potential. However, our splitting of the velocity field into solenoidal and irrotational parts compensates for this handicap to the extent that we obtain results of similar quality to those of \cite{Uryu:1999uu} with modest grid resolutions (from 20 to 30 grid points across the star).

\begin{figure}
\epsfxsize=2.8in
\begin{center}
\leavevmode \epsffile{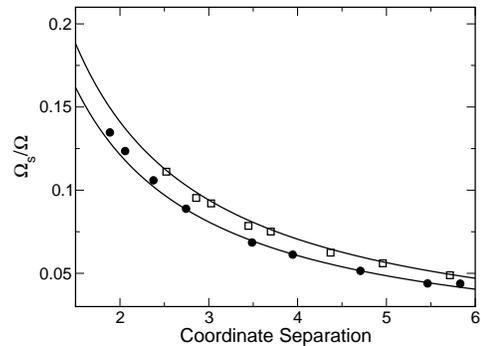}
\end{center}
\caption{ Ratio of spin vs orbital angular velocities as a function of the coordinate separation. The filled circles (open squares) correspond to the $(m/R)_\infty = 0.14 ~(0.19)$ and ${\cal C}=0$ sequence. The solid lines show the PPN prediction (Appendix \ref{appendix_circ}).}
\vspace {4 mm}
\label{Spin_vs_d}
\end{figure}

An interesting effect is revealed on the plot of the spin parameter $a= \Omega_s/\Omega$ versus coordinate separation between stellar centers. Figure \ref{Spin_vs_d} shows the spin parameter for the ${\cal C} =0$ sequences for both compaction ratios $(m/R)_\infty = 0.14$ and $(m/R)_\infty = 0.19$. We see that the stellar spins, which point along the direction of the orbital angular momentum, increase as the stars get nearer. For comparison we also show the corresponding post-Newtonian prediction. As it can be seen, the agreement is close all the way up to the closest orbits (see Appendix \ref{appendix_circ} for more details and post-Newtonian derivation).


\subsection{Constant circulation sequences}
\label{const_circulation_seq}

\begin{figure}
\epsfxsize=2.8in
\begin{center}
\leavevmode \epsffile{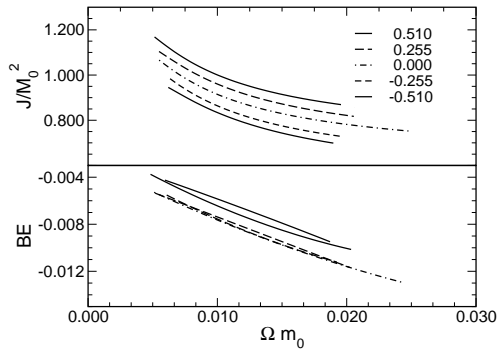}
\end{center}
\caption{ Angular momentum (upper) and binding energy (lower) vs orbital angular frequency for the sequences with $(m/R)_\infty=0.14$. The lines are polynomial fits to the data and the labels denote the circulation ratio ${\cal C} /  {\cal C}_{crit}$ for the curves from top to bottom (upper plot).}
\vspace{2 mm}
\label{CR_0.14_J_BE}
\end{figure}

We constructed five sequences for the $(m/R)_\infty = 0.14$ binary, with circulations ${\cal C} /  {\cal C}_{crit} = (-0.510,-0.255, ~0.0, ~0.255, ~0.510)$, and three  for the $(m/R)_\infty = 0.19$ system with circulations ${\cal C} /  {\cal C}_{crit} = (-0.124, ~0.0, ~0.247)$. 

Figures \ref{CR_0.14_J_BE} and \ref{CR_0.19_J_BE} show the angular momentum and binding energy as a function of the orbital frequency for the cases $(m/R)_\infty = 0.14$ and $(m/R)_\infty = 0.19$, respectively. As expected, the angular momentum increases with the circulation ${\cal C}$, due to the addition (or subtraction, when ${\cal C} < 0$) of spin angular momentum to the system. In the same way, the binding energy increases with $|{\cal C}|$ due to the extra rotational kinetic energy. Note that none of these curves show turning points prior to termination, in contrast to corotating sequences \cite{bcsst}. With the exception of the $(m/R)_\infty = 0.14$ and ${\cal C} = 0.0$ sequence (Fig.  \ref{Cont_rho_cusp}), we did not continue the orbits all the way up to the point of contact for two reasons: first, the code requires very high resolution to resolve the cusp at the contact point, making the calculation quite resource-intensive. Second, the closest orbits treated here appear to be inside the ISCO, thus rendering any analysis of the physical aspects of such orbits meaningless \cite{footnote8}. This claim is based on the dynamical determination of the ISCO that will be reported in a forthcoming paper \cite{Marronetti:2003aa}, in which the initial data sets presented here are evolved for a couple of orbits. Also, due to the high resolution and large CPU time requirements, only three sequences are presented for the case $(m/R)_\infty = 0.19$. Future work contemplates a more extensive coverage of the range of compaction ratios and circulations \cite{Marronetti:2003bb}.

\begin{figure}
\epsfxsize=2.8in
\begin{center}
\leavevmode \epsffile{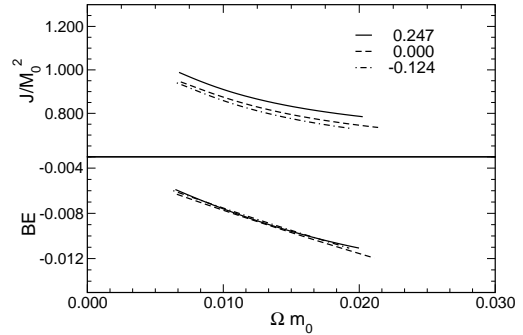}
\end{center}
\caption{ Angular momentum (upper) and binding energy (lower) vs orbital angular frequency for the sequences with $(m/R)_\infty=0.19$.  The labels denote the circulation ratio ${\cal C} /  {\cal C}_{crit}$ for the curves from top to bottom (upper plot).}
\label{CR_0.19_J_BE}
\end{figure}

Figure \ref{CR_Os_vs_Ob_Full} shows the spin vs orbital angular velocity for both sets of sequences, while Fig. \ref{CR_Os_vs_Ob_0.0} compares only the ${\cal C}=0$ sequences for both compaction ratio cases. The neutron stars spin up during the inspiral, starting from zero spin at infinity to spin frequencies that are $13\%$ ($11\%$) of the orbital angular frequencies for the sequence with $(m/R)_\infty = 0.14$ ($0.19$). This spin-up effect is observed in all the sequences with the exception of the $(m/R)_\infty = 0.14$ and ${\cal C} = 0.510$ case (the largest circulation value studied in this paper), where the spin rate decreases. This spin-up effect is imprinted in the inspiral gravitational radiation waveform and potentially observable in the electromagnetic spectrum if one of the neutron stars is also a pulsar (radio or X-ray).

Figure \ref{CR_a_vs_d} shows the ratio of spin angular velocity vs orbital angular velocity as a function of the coordinate separation.

\begin{figure}
\epsfxsize=2.8in
\begin{center}
\leavevmode \epsffile{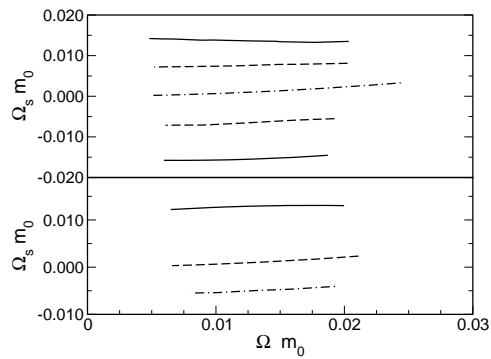}
\end{center}
\caption{ Spin vs orbital angular velocities for the sequences with $(m/R)_\infty=0.14$ (upper) and $(m/R)_\infty=0.19$ (lower). The line captions are those of Fig. \ref{CR_0.14_J_BE} (upper) and Fig. \ref{CR_0.19_J_BE} (lower).}
\label{CR_Os_vs_Ob_Full}
\end{figure}

\begin{figure}
\epsfxsize=2.8in
\begin{center}
\leavevmode \epsffile{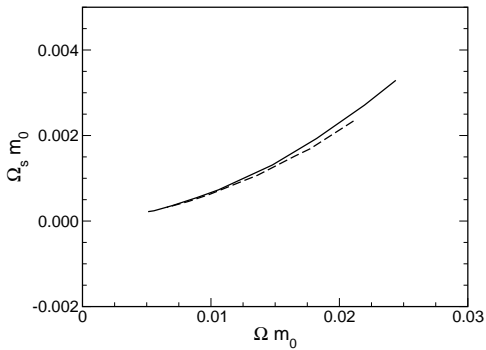}
\end{center}
\caption{ Spin vs orbital angular velocities for the ${\cal C}=0$ circulation sequence for compaction ratios $(m/R)_\infty=0.14$ (solid line) and $(m/R)_\infty=0.19$ (dashed line).}
\label{CR_Os_vs_Ob_0.0}
\end{figure}

\begin{figure}
\epsfxsize=2.8in
\begin{center}
\leavevmode \epsffile{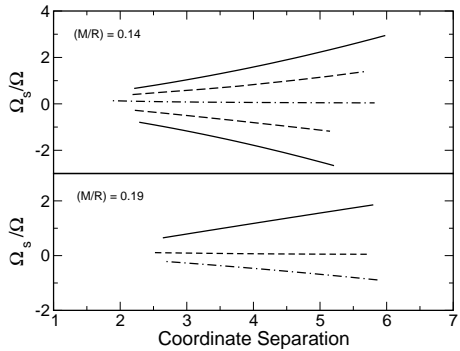}
\end{center}
\caption{ Ratio of spin vs orbital angular velocities for the sequences with $(m/R)_\infty=0.14$ (upper) and $(m/R)_\infty=0.19$ (lower). The line captions are those of Fig. \ref{CR_0.14_J_BE} (upper) and Fig. \ref{CR_0.19_J_BE} (lower).}
\label{CR_a_vs_d}
\end{figure}


\section{Conclusions}
\label{conclusions}

We introduce a new method for constructing quasi-equilibrium sequences of binary neutron stars in circular orbit, allowing for the free specification of the rotational component of the fluid velocity. This allows us to choose the spin of the stars, thus extending the traditional options of corotating and irrotational binaries. The sequences of circular orbits are characterized by two free parameters: the baryonic mass of each star, as in the standard quasi-equilibrium approach, and the relativistic equatorial circulation ${\cal C}$. The conservation of the latter during inspiral along hypersurfaces of constant proper time is guaranteed by the Kelvin-Helmoltz theorem \cite{c79}. Additionally, the numerical formalism is designed to produce high quality initial data sets using Cartesian grids with modest resolution for future evolution studies.

We constructed sequences for different values of the circulation for binaries with stars of moderate and high compaction ratio. Each circulation value corresponds to a particular value of spin (perpendicular to the orbital plane) when the stars are at infinite separation. These sequences show the spin-orbital angular momentum coupling that accompanies the inspiral (in the framework of the quasi-equilibrium binary formalism). In particular, we find that the nonlinear gravitational fields induce a spin-up in the stars for all the cases, except the sequence with the largest value of ${\cal C}$. The variation of the spin of the star throughout the inspiral is a potentially observable electromagnetic effect when one of the neutron star is a pulsar.

We plan a more extensive study of the parameter space ($M_0 / {\cal C}$) for different EOSs and a detailed study of the behavior of the central density vs separation throughout the inspiral \cite{Marronetti:2003bb}. The quasi-equilibrium orbits presented here are particularly well suited as initial data sets for evolutionary codes. In a forthcoming paper, we will study the evolution of such a set for binaries at different separations tracking them for the last couple of orbits to determine dynamically the location of the ISCO \cite{Marronetti:2003aa}.


\acknowledgments
It is a pleasure to thank Yuk-Tung Liu, Hwei-Jang Yo, Matt Duez, and Eric Gourgoulhon for helpful discussions. This paper was supported in part by NSF Grants PHY-0090310 and PHY-0205155 and NASA Grant NAG 5-10781 at UIUC. PM gratefully acknowledges financial support through the Fortner Fellowship at UIUC. This work was partially supported by National Computational Science Alliance under Grants PHY020008N and MCA99S008, and utilized the IBM Regatta p690 computer cluster.


\begin{appendix}

\section{Continuity equation in conformally flat space}
\label{appendix_Cont}

The continuity equation (\ref{cont1}) expresses the conservation of baryonic mass in the system. In this appendix we will derive the expression for such equation in the rotating frame when the spatial metric tensor is conformally flat. 

The continuity equation can be split as
\ba \label{ap_1}
(\rho_0 u_R^\mu)_{;\mu} = \nabla_\mu (\rho_0 u_R^\mu) + \rho_0 ~\Gamma^\mu_{\mu\nu} u_R^\nu = 0~.
\ea
Since the system is stationary in the rotating frame, any partial time derivatives vanish, reducing the first term of the RHS of Eq. (\ref{ap_1}) to
\ba \label{first_term}
\nabla_\mu (\rho_0 u_R^\mu) = \nabla_0 (\rho_0 u_R^0) + \nabla_l (\rho_0 u_R^l) = \nabla_l (\rho_0 u_R^l)~.
\ea
We concentrate now on the term $\rho_0 ~\Gamma^\mu_{\mu\nu} u_R^\nu$. Using the definition of the affine connection, we see that the contraction $\Gamma^\mu_{\mu 0}$ vanishes in the rotating frame
\ba
\Gamma^\mu_{\mu 0} &=& \frac{1}{2} g^{\mu\sigma} (\nabla_0 g_{\mu\sigma}+\nabla_\mu g_{\sigma 0}-\nabla_\sigma g_{\mu 0}) \nonumber \\
&=& \frac{1}{2} g^{\mu\sigma} (\nabla_\mu g_{\sigma 0}-\nabla_\sigma g_{\mu 0}) = 0 ~. \nonumber
\ea
The term $\rho_0 ~\Gamma^\mu_{\mu\nu} u_R^\nu$  gets reduced to the sum over the spatial components of the four-velocity $u_R^\nu$
\ba \label{second_term}
\rho_0 ~\Gamma^\mu_{\mu\nu} u_R^\nu =\rho_0 ~\Gamma^\mu_{\mu l} u_R^l~.
\ea
In order to evaluate the sum over the affine connection components $\Gamma^\mu_{\mu l}$, we will use the conformally flat spatial metric introduced by Eq. (\ref{conftensor}), together with the relations between the lapse, shift vector, and spatial metric and the components of the four dimensional metric tensor $g_{\mu\nu}$ \cite{MTW}. After some algebraic steps, we get
\ba
\Gamma^0_{0l} & = & \frac{1}{\alpha} ~\nabla_l \alpha - \frac{1}{2\alpha^2} ~[ \Psi^4 ~(\beta^m \nabla_l \beta^m+\beta^m \nabla_m \beta^l) \nonumber \\
& + & 4 \Psi^3 \beta^l \beta^m ~\nabla_m \Psi ] \nonumber \\
\Gamma^m_{ml} & = & \frac{6}{\Psi} \nabla_l \Psi + \frac{1}{2\alpha^2} ~[ \Psi^4 ~(\beta^m \nabla_l \beta^m+\beta^m \nabla_m \beta^l) \nonumber \\
& + & 4 \Psi^3 \beta^l \beta^m ~\nabla_m \Psi ]~, \nonumber
\ea
where the raising (lowering) of indices is done using $\delta^{ij}$ ($\delta_{ij}$) which we have omitted for simplicity. With this result, we write Eq. (\ref{second_term}) as
\ba \label{second_term2}
\rho_0 ~\Gamma^\mu_{\mu\nu} u_R^\mu & = & \rho_0 ~\Gamma^\mu_{\mu l} u_R^l \nonumber \\
& = & \rho_0 ~(\Gamma^0_{0 l}+\Gamma^m_{ml}) u_R^l \nonumber \\
& = & \rho_0 u_R^l ~\left( \frac{1}{\alpha} \nabla_l \alpha + \frac{6}{\Psi} \nabla_l \Psi \right) \nonumber \\
& = & \rho_0 u_R^l ~\nabla_l \ln(\alpha \Psi^6) ~.
\ea
Replacing Eqs. (\ref{first_term}) and (\ref{second_term2}) in Eq. (\ref{ap_1}), we get
\ba
\nabla_l(\rho_0 u^l_{R}) + \rho_0 u^l_R \nabla_l[\ln(\alpha \Psi^6)] = 0~, \nonumber
\ea
which we divide by $u^0$
\ba
\frac{1}{u^0} \nabla_l(\rho_0 u^l_R) + \rho_0 \frac{u^l_R}{u^0} \nabla_l[\ln(\alpha \Psi^6)] = 0 ~.
\label{cont2}
\ea
The first term can be decomposed
\ba
\frac{1}{u^0} \nabla_l(\rho_0 u^l_R) & = & \nabla_l \left( \rho_0 \frac{u^l_R}{u^0} \right)-\rho_0 u^l_R  \nabla_l \left( \frac{1}{u^0} \right) \nonumber \\
& = & \nabla_l \left(\rho_0 \frac{u^l_R}{u^0} \right) + \rho_0 \frac{u^l_R}{u^0} \nabla_l(\ln(u^0)), \nonumber
\ea
which, when combined with the second term, leads to a new expression for the continuity equation (\ref{cont2})
\ba
\nabla_l \left(\rho_0 \frac{u^l_R}{u^0} \right) + \rho_0 \frac{u^l_R}{u^0} \nabla_l(\ln(u^0 \alpha \Psi^6)) = 0~. \nonumber
\ea
For convenience, we will use the fluid three-velocity $v^i_R \equiv {u^i_R}/{u^0}$ as our variable, thus arriving at the expression (\ref{cont3})
\ba
\nabla_l(\rho_0 v^l_R) + \rho_0 v^l_R \nabla_l(\ln(u^0 \alpha \Psi^6)) = 0~. \nonumber
\ea


\section{Bernoulli equation for Corotating binaries}
\label{appendix_Bern}

Setting $a=1$ in Eq. (\ref{vI}) gives the three-dimensional coordinate velocity in the inertial $v^i_I = (\vec{\Omega} \times \vec{r})^i$, which is identical to the spatial components of the Killing vector $\xi_I^\mu$ in the same frame. We can write the Killing vector as
\ba
\xi_I^\mu = (1, v^i_I)=(1,u^i_I/u^0)~. \nonumber
\ea
As expected, the Killing vector is proportional to the fluid four-velocity: $\xi_I^\mu = u^\mu_I/u^0$. The field $F$ becomes
\ba
F \equiv u_{I\mu} \xi_I^\mu = -\frac{1}{u^0}~, \nonumber
\ea
where we have used the normalization condition (\ref{norm_vel}). Using this expression together with Eq. (\ref{bern3}), we get 
\ba
\frac{u^0}{h} = const~, \nonumber
\ea
which is the familiar expression of the Bernoulli equation for corotating binaries with polytropic EOS (see for instance Eq. (32) of \cite{bcsst}).


\section{Newtonian Limits}
\label{appendix_Newt}

We derive in this appendix the Newtonian limit of the hydrodynamical equations employed in our method. The gravitational part is simply given by Poisson's equation for the Newtonian potential $\phi_N$. We start by keeping only the terms to first order in the Newtonian potential $\phi_N$, the Newtonian enthalpy $h_N$, the rest mass density $\rho_0$, the square of the fluid velocity in the inertial frame $v_I^2$, and the contraction $\delta_{lm}(\vec{\Omega} \times \vec{r})^l v_I^m$. The simplest limits are
\ba
h        & \rightarrow & 1 + h_N \nonumber \\
\alpha   & \rightarrow & 1 + \phi_N \nonumber \\
\beta^i  & \rightarrow &  (\vec{\Omega} \times \vec{r})^i~, \nonumber
\ea
These limits in combination with the equation for the time component of the fluid four-velocity $u^0$ (\ref{u0}) give
\ba \label{u0_lim}
u^0 \rightarrow & 1 - \phi_N + v_I^2/2~,
\ea
which together with $\xi_I^i = (\vec{\Omega} \times \vec{r})^i$, yield the limit of the field $F$ (\ref{F3})
\ba
F \rightarrow & - [1 + \phi_N + v_I^2/2 - \delta_{lm}(\vec{\Omega} \times \vec{r})^l v_I^m]~. \nonumber
\ea
Now the limit of the Bernoulli integral (\ref{bern3}) is simply
\ba
1 + \phi_N + v_I^2/2 & - & \delta_{lm}(\vec{\Omega} \times \vec{r})^l v_I^m + h_N = C'~. \nonumber
\ea
Making $\tilde{C} = C'-1$ a new constant, we get the Newtonian Bernoulli equation
\ba \label{bern_lim}
\phi_N + v_I^2/2 & - & \delta_{lm}(\vec{\Omega} \times \vec{r})^l v_I^m + h_N = \tilde{C}~.
\ea

The Newtonian limit of the elliptic equation for the potential $\sigma$ is easy to obtain by realizing that the limit of the argument of the logarithm in the RHS of Eq. (\ref{sigma_eq}) is just $\rho_0$. Then, the Newtonian limit of Eq. (\ref{sigma_eq}) becomes
\ba \label{sigma_lim}
\nabla^2 \sigma = -(v^l_{RS}+\nabla^l \sigma) \frac{\nabla_l (\rho_0)}{\rho_0} ~.
\ea
In similar manner, we get the limit for the elliptic equation for the flow potential $\lambda$
\ba \label{lambda_lim}
\nabla^2 \lambda = \nabla_l(\lambda ~\nabla^l \ln(\rho_0)) - v^l_{RS} ~\nabla_l \rho_0 ~.
\ea

Equations (\ref{bern_lim}) and (\ref{sigma_lim}) [or equivalently Eq. (\ref{lambda_lim})], together with Poisson's equation for the Newtonian potential $\phi_N$ form the complete self-consistent set of equations necessary for applying our method in Newtonian gravitation, appropriate for scenarios like binary white dwarfs, etc. Note that in the case ${\cal C}=0$ and $a=0$, Eqs. (\ref{bern_lim}) and (\ref{sigma_lim}) reduce to the equations derived by Bonazzola {\it et al.} \cite{Bonazzola:1992} for Newtonian irrotational binaries.


\section{Calculation of Masses, Angular Momentum, and Circulation}
\label{appendix_int}

Every quasi-stable circular orbit in each sequence is characterized by a set of global quantities: the rest mass $M_0$, the gravitational or ADM mass $M$, the total angular momentum $J$, and the circulation along the stellar equator $\cal{C}$. In this appendix we present the formulas used to evaluate those quantities using the ``$\kappa = 1$" unit system described in Sec. \ref{set_equations}.

The rest mass $M_0$ of the system is \cite{Duez:2002bn}
\ba
M_0 = \int_{\cal{V}} \Psi^6 \alpha u^0 q^n d^3x~, \nonumber
\ea
where we have used $\rho_0=q^n$. The total gravitational mass of the system can be read off of the monopole term of the conformal factor $\Psi$, given that $\Psi = 1 + M/2r + \mathcal{O}(r^{-2})$ \cite{Wilson:1996ty}
\ba
M = - \frac{1}{2 \pi} \oint_{\infty} \nabla^i \Psi dS_i = - \frac{1}{2 \pi} \int_{\infty} \nabla^2 \Psi d^3x~. \nonumber
\ea 
This last volume integral is evaluated replacing $\nabla^2 \Psi$ by the RHS of the first (Hamiltonian) equation of the set (\ref{final_eqs}).

The orbital angular momentum of the binary is (by construction) positive along the $\hat{z}$ direction. Since we only consider stars with spins along the $z$ axis, the total ADM angular momentum of the system remains confined to this direction. Following
\cite{by80} we write
\ba
J = J_z &=& \frac{\epsilon_{zjk}}{8 \pi} \oint_\infty x^j \tilde{K}^{lk}~dS_l \nonumber \\
&=& \frac{\epsilon_{zjk}}{8 \pi} \int_\infty x^j \nabla_l \tilde{K}^{lk} d^3x~. \nonumber
\ea
Using the identity $\nabla_l \tilde{K}^{lk} = \Psi^{10} D_l K^{lk}$ and the momentum constraint (\ref{constraints}) we get
\ba
J &=& \int_\infty \Psi^{10} ~(x~j^y - y~j^x) ~d^3x \nonumber\\
&=& \int_\infty \Psi^{10} ~q^n ~[1+(1+n)q] ~{u^0}^2 ~\alpha ~ \nonumber \\
& & [x~(v_R^y+\beta^y) - y~(v_R^x+\beta^x)] ~d^3x ~, \nonumber
\ea
where we have used some of the expressions derived in Sec. \ref{EOS}.

The relativistic circulation $\cal{C}$, defined by Eq. (\ref{circ1}) presents the numerical complication of requiring a line integral along the equator which, being a quasi-circular path, makes the result sharply dependent on the resolution of our Cartesian grid. To minimize this effect, we transform the line integral into a surface integral using Stokes theorem
\ba \label{stokes}
\int_{\Omega} d\tilde{\Theta} = \int_{\partial \Omega} \tilde{\Theta}~.
\ea
In the case of Eq. (\ref{circ1}), we have
\ba
\tilde{\Theta} = h u_{\mu} d\tilde{x}^\mu  ~, \nonumber
\ea
which gives
\ba
d\tilde{\Theta} & = & d(h~u_\mu) \wedge d\tilde{x}^\mu + h ~ u_\mu \wedge d(d\tilde{x}^\mu) \nonumber \\
& = & (h~u_\mu)_{,\nu} ~d\tilde{x}^\nu \wedge d\tilde{x}^\mu~. \nonumber
\ea
Using Eq. (\ref{stokes}), we can write the circulation as
\ba
{\cal C}(c) = \oint_{c} h u_{\mu} d\sigma^{\mu} = \int_S [(h~u_y)_x - (h~u_x)_y] dS~, \nonumber
\ea
where $S$ is the surface confined by the equator on the $z=0$ plane.


\section{Elliptic Solver Iterative Algorithm}
\label{appendix_solver}

The set of elliptic equations (\ref{final_eqs}) and the Bernoulli equation (\ref{bern4}) are solved iteratively, starting with an initial guess for the fields. The iteration process is illustrated by the following pseudo-code:

\vspace{0.5cm}
{\bf Set Binary parameters} : $m_0 ~(=M_0/2)$, $d$, ${\cal{C}}_\infty$

{\bf Set Initial Iteration Parameters} : $\Omega_0$, $(\delta \Omega)_0$

{\bf Set Fields Initial Guess} 

\vspace{0.2cm}
{\bf Do} $iter$
\vspace{0.1cm}

~~{\bf Do} $index=1,7$

~~~~{\bf Solve}($\nabla^2 F^{index}_{iter} = \rho^{index}_{iter}$)

~~{\bf End Do} $index$
\vspace{0.1cm}

~~{\bf Do} $j=-1,1$

~~~~$\tilde{\Omega}_j=\Omega_{iter-1} + j ~(\delta \Omega)_{iter-1}$

~~~~{\bf Compute} $\tilde{q}_{j}$ {\bf using} $\tilde{\Omega}_j$

~~~~{\bf Calculate} $(\Delta q)_j = ||\tilde{q}_{j} - q_{iter-1}||_2$

~~{\bf End Do} $j$
\vspace{0.1cm}

~~$J = j$, such that $(\Delta q)_j = \min((\Delta q)_{-1},(\Delta q)_0,(\Delta q)_1)$

~~{\bf Set} $q_{iter} = \tilde{q}_{J}$

~~{\bf Set} $\Omega_{iter} = \tilde{\Omega}_J$

~~{\bf If} $(J=0)$ {\bf then}

~~~~$(\delta \Omega)_{iter} = (\delta \Omega)_{iter-1}/2$

~~{\bf else}

~~~~$(\delta \Omega)_{iter} = (\delta \Omega)_{iter-1}$

~~{\bf End If}
\vspace{0.1cm}

~~{\bf Find} $C$ such that the rest mass is $m_0$
\vspace{0.1cm}

~~{\bf Find} $a$ such that the circulation is ${\cal{C}}_\infty$
\vspace{0.1cm}

~~{\bf Compute} $(\Delta F)^{index} \equiv ||F^{index}_{iter} - F^{index}_{iter-1}||_2$

~~{\bf If} $(\Delta F)^{index} < \epsilon$) for ($1 \leq index \leq 7$) {\bf then STOP}

\vspace{0.1cm}
{\bf End Do} $iter$\\

\vspace{0.5cm}
The following is a brief description of the main features of each code section:
\begin{description}
\item[Set Binary Parameters:] The three free parameters of an identical star binary system are set. For our code we have chosen the individual star rest mass $m_0$, the coordinate separation distance $d$, and the equatorial circulation at infinite separation ${\cal{C}}_\infty$.
\item[Set Initial Iteration Parameters:] We start the iteration providing an initial value to the binary angular velocity $\Omega_0$ and its error bracket $(\delta \Omega)_0$. This last value will be reduced during the iteration process, until it reaches the desired accuracy.
\item[Set Fields Initial Guess:] The gravitational fields and the rest mass density $q$ are initially set by mapping the solution of the TOV equations onto the octant grid.
\item[Loop $iter$:] Main relaxation loop of the code.
\item[Loop $index$:] This loops goes across the list of seven elliptic equations. The source term and boundary conditions of equation $index$ are computed using the values of fields 1 to $(index-1)$ at iteration step $iter$ and the values of fields $index$ to 7 at iteration step $(iter-1)$.
\item[Loop $j$:] This loops determines the next value of the angular velocity $\Omega$. The rest-mass density field $q_j$ is computed using the Bernoulli equation (\ref{bern4}) for three different values of the angular velocity $\tilde{\Omega}_j$, namely ($\Omega_{(iter-1)}-(\delta \Omega)_{(iter-1)}$), $\Omega_{(iter-1)}$, and ($\Omega_{(iter-1)}+(\delta \Omega)_{(iter-1)}$). The value of $\Omega_{iter}$ is set to the value of $\tilde{\Omega}_j$ that minimizes the $L_2$ norm of the difference between $q_j$ and $q_{(iter-1)}$ and the new density profile $q_{iter}$ is set to this field $q_j$. If the value selected for the angular velocity $\Omega_{iter}$ is the corresponding to $j=0$, then the value of $(\delta \Omega)_{iter}$ is set to half of $(\delta \Omega)_{(iter-1)}$, otherwise it is left unchanged.
\item[Rest Mass:] The Bernoulli constant $C$ from Eq. (\ref{bern4}) is adjusted to keep the desired value $m_0$.
\item[Circulation:] The spin parameter $a$ is adjusted to keep the circulation at the desired value ${\cal{C}}_\infty$.
\item[Tolerance Calculation:] The new fields are compared with their respective values at the previous iteration step by means of the $L_2$ norm. If the norm of the difference falls below some threshold value $\epsilon$ for all the fields (i.e., $index$ from 1 to 7), then the iteration is stopped.
\end{description} 


\section{Calculation of the Circulation using PPN expansions}
\label{appendix_circ}

Using parametrized post-Newtonian expansions of the Einstein field equations we
can estimate the relativistic circulation of fluid $\cal C$ around a closed path $c$ inside the neutron star. We will consider the case of a binary in the inertial frame located at the system's center of mass. In this appendix, $u^\mu$ and $\beta^i$ will represent the fluid four-velocity and shift vector in the inertial frame. The binary is composed by a neutron star with parameters $m_s$, $\vec{V}_s$, and $\vec{r}_s$, and a companion body with parameters  $m_b$, $\vec{V}_b$, and $\vec{r}_b$. These parameters are the mass, orbital velocity, and coordinate position of the stellar center respectively. Additionally, we will assume that the velocity of the fluid behaves like a rigid rotator 
\ba
\vec{V} = \vec{V}_s + [\vec{\Omega}_s \times (\vec{r}-\vec{r}_s)]^i ~, \nonumber
\ea
where $\vec{\Omega}_s$ is the spin angular velocity of the star. Note that we are ignoring any irrotational components of the velocity, like the term $\vec{\nabla} \sigma$ in Eq. (\ref{vI}) since, as it will be clear below, only the solenoidal part of the velocity will survive.

To estimate the magnitude and scaling behavior of the fluid velocity we will take the components of the four-metric to be given by the series expansion
\ba \label{ppn_metric}
g_{00} &=& -1-2 ~\phi_N + {\cal O}(\epsilon^2) \nonumber \\
g_{0i} &=& \beta_i = -4 ~\chi_i + {\cal O}(\epsilon^{5/2}) \nonumber \\
g_{ij} &=& \gamma_{ij} = (1-2 ~\phi_N) ~\delta_{ij} + {\cal O}(\epsilon^{2})~.
\ea
Here $\epsilon$ is the PPN expansion parameter ($\epsilon \sim M/r \sim v^2$) and the Newtonian potential $\phi_N$ and the potential $\chi_i$ are adapted from the expressions for the near-zone for point masses given by Will and Wiseman \cite{Will:1996} 
\ba \label{ppn_pot}
\phi_N &=& \phi_s(\vec{r})- \frac{m_b}{|\vec{r}-\vec{r}_b|} \nonumber \\
\chi_i &=&  -\phi_s(\vec{r}) V^i_s + \frac{m_b V^i_b}{|\vec{r}-\vec{r}_b|}~,
\ea
where $\phi_s(\vec{r})$ is the self-gravitating Newtonian potential of the primary neutron star. We will ignore the effect of the spin of the bodies in the shift vector $\beta^i$, since it will only contribute with terms that appear at higher order in $\epsilon$.

The relativistic circulation is given by Eq. (\ref{circ1})
\ba  \label{ap_circ}
{\cal C}(c) &=& \oint_{c} h u_{\mu} d\sigma^{\mu} \nonumber\\
& = & \oint_{c} h u_i dx^i = \oint_{c} h \vec{v} \cdot d\vec{x} = \int_{S(c)} \vec{\nabla} \times (h~\vec{v})  \cdot d\vec{S} \nonumber\\
& = & \int_{S(c)} [h ~(\vec{\nabla} \times \vec{v}) + \vec{\nabla}h \times \vec{v}]  \cdot d\vec{S}~,
\ea
where we have used the fact that the line element along the closed path $c$ is purely spatial $d\sigma^\mu = (0,dx^i)$, and Stokes theorem. The three-dimensional vector $\vec{v}$ has been defined by the spatial components $u_i$. We evaluate those components using Eq. (\ref{u0_lim}) and the metric (\ref{ppn_metric})
\ba \label{u_vec}
(\vec{v})_i & \equiv & u_i = \beta_i u^0 + \gamma_{ij} u^j \nonumber \\
&=& \beta_i + (1 - 3 ~\phi_N) V^i + {\cal O}(\epsilon^2)~, \nonumber
\ea
where $V^i = u^i/u^0$. We then obtain
\ba
\vec{v} \simeq \vec{\beta} + \vec{V} - 3 ~\phi_N \vec{V}~.
\ea
With the definition of the PPN potentials (\ref{ppn_pot}), we obtain the following expressions for the curl of $\vec{v}$ (\ref{u_vec})
\ba \label{u_vec2}
\vec{\nabla} \times \vec{\beta} & = & 4 \vec{\nabla} \phi_s \times \vec{V}_s + \frac{4 ~m_b}{|\vec{r}-\vec{r}_b|^3} (\vec{r}-\vec{r}_b) \times \vec{V}_b \nonumber \\
\vec{\nabla} \times \vec{V} & = & 2 \vec{\Omega}_s~,
\ea
and splitting the last term of the RHS of Eq. (\ref{u_vec})
\ba
\phi_N \vec{V} & = & \phi_N \vec{V}_s + \phi_N ~[\vec{\Omega}_s \times (\vec{r}-\vec{r}_s)] \nonumber \\
& = & \phi_s \vec{V}_s - \frac{m_b ~\vec{V}_s}{|\vec{r}-\vec{r}_b|} + \phi_N [\vec{\Omega}_s \times (\vec{r}-\vec{r}_s)] ~,\nonumber
\ea
we get
\ba \label{curl_3}
-3 \vec{\nabla} \times (\phi_N\vec{V}) = &-& 3 \vec{\nabla} \phi_s(\vec{r}) \times \vec{V}_s \nonumber \\
& - & \frac{3 ~m_b}{|\vec{r}-\vec{r}_b|^3} (\vec{r}-\vec{r}_b) \times \vec{V}_s - 6 \phi_N \vec{\Omega}_s  \nonumber \\
& - & 3 \vec{\nabla} \phi_N \times [\vec{\Omega}_s \times (\vec{r}-\vec{r}_s)] ~.
\ea

Let us now consider the particular case of an irrotational fluid. By definition, the vorticity of the fluid is null and thus the circulation of the fluid along {\it any} closed path $c$ inside the star will be zero. Equation (\ref{ap_circ}) shows that this is only satisfied if 
\ba \label{integrand1}
h ~(\vec{\nabla} \times \vec{v}) + \vec{\nabla}h \times \vec{v} = 0
\ea
everywhere in the stellar volume, in particular the stellar center $\vec{r}=\vec{r}_s$. The center is defined as the point of maximum density (or equivalently enthalpy), then $\vec{\nabla} h (\vec{r}=\vec{r}_s)=0$. Since $h > 0$,  at the stellar center Eq. (\ref{integrand1}) becomes $\vec{\nabla} \times \vec{v}=0$.
An expression for $\vec{\nabla} \times \vec{v}$ can be constructed from Eqs. (\ref{u_vec}), (\ref{u_vec2}),  (\ref{curl_3}). Equating that expression to zero, setting $\vec{r}=\vec{r}_s$, and solving for $\vec{\Omega}_s$ gives 
\ba \label{om_2}
\vec{\Omega}_s |_{\vec{r}=\vec{r}_s} &=& \frac{m_b}{2 |\vec{r}_s-\vec{r}_b|^3} (\vec{r}_s-\vec{r}_b) \times (3 ~\vec{V}_s - 4 ~\vec{V}_b) \nonumber \\
& & (1 - 3~\phi_N(\vec{r}_s))^{-1} \nonumber \\
&\simeq& \frac{m_b}{2 |\vec{r}_s-\vec{r}_b|^3} (\vec{r}_s-\vec{r}_b) \times (3 ~\vec{V}_s - 4 ~\vec{V}_b) ~,
\ea
where we use the fact that $\vec{\nabla} \phi_s (\vec{r}=\vec{r}_s)=0$ since $\phi_s (\vec{r}) \propto (\vec{r}-\vec{r}_s)^2$ in the neighborhood or $\vec{r}_s$. Note that the self-gravity terms in the metric cancel out of the final expression. Since we are working in the center of mass system and the objects are in circular orbit around the origin of coordinates, we have 
\ba \label{relations}
\vec{r}_s &=& \frac{m_b}{M} ~d ~\hat{r}_s \nonumber \\
\vec{r}_b &=& -\frac{m_s}{M} ~d ~\hat{r}_s \nonumber \\
\vec{V}_s &=& \vec{\Omega} \times \vec{r}_s \nonumber \\
\vec{V}_b &=& -\frac{m_s}{m_b} ~\vec{V}_s~,
\ea
where $d \equiv |\vec{r}_s-\vec{r}_b|$, $\hat{r}_s \equiv \vec{r}_s / r_s$, $M \equiv (m_s+m_b)$, and $\vec{\Omega}$ is the orbital angular velocity of the binary. Inserting the relations (\ref{relations}) into Eq. (\ref{om_2}) gives
\ba \label{om_3}
\vec{\Omega}_s &=& \frac{m_b ~(3 m_b+ 4 m_s)}{2 ~M ~d} ~\vec{\Omega} \nonumber \\
&=& \frac{3 m_b + \mu}{2 ~d} ~\vec{\Omega}~,
\ea
where $\mu \equiv (m_b ~m_s)/(m_s+m_b)$ is the reduced mass. In the case of equal mass binaries like the ones studied on this paper, Eq. (\ref{om_3}) reduces to
\ba \label{membr_om}
\vec{\Omega}_s &=& \frac{7}{4} \frac{m_s}{d}~\vec{\Omega} \nonumber~.
\ea
Note that Eq. (\ref{om_3}) is identical (to PPN order) to the formula for the precession of the spin of a star with respect to the inertial frame tied to distant galaxies. This corresponds to a combination of the Lense-Thirring (gravitomagnetic) and the geodetic effects \cite{Membrane}.

The PPN lines shown on Fig. \ref{Spin_vs_d} were calculated as $\Omega_s/\Omega = \frac{7}{4} ~m/d$, where $m$ is the gravitational mass of a star in isolation.
 

\section{Tables}
\label{Tables}

This appendix summarizes the parameters corresponding to each circular orbit presented in this paper. Tables III-X group the orbits belonging to a sequence of identical $\Gamma = 2$ binary polytropes with given compaction ratio $(m/R)_\infty$ and relativistic equatorial circulation ${\cal C}$. The parameters tabulated are the coordinate separation $d$, the total gravitational mass $M$, the binding energy $BE \equiv (M - M_\infty) / M_0$, the total angular momentum $J / M_0^2$, the orbital angular frequency $\Omega ~m_0$, and the spin angular momentum $\Omega_s ~m_0$. The total rest mass of the system with $(m/R)_\infty=0.14$ ($0.19$) is $M_0=0.2938$ ($0.3534$). The critical value of equatorial circulation which corresponds to the case of single stars rotating at the mass-shedding limit is ${\cal C}_{crit}=1.96$ ($2.02$) for the compaction ratio $(m/R)_\infty=0.14$ ($0.19$).

\begin{table*}[h]
\begin{center}
\caption{$(m/R)_{\infty} = 0.140$, ${\cal C}/{\cal C}_{crit} = -0.510$.}
\begin{tabular}{cccccc}
\hline
\hline
~~$d$~~&~~$M$~~&~~$BE$~~&~~$J/M_0^2$~~&~~$\Omega ~m_0$~~&~~$\Omega_s ~m_0$~~\\
\hline
  2.29 ~~&~~ 0.2709 ~~&~~ -9.49 $\times~10^{-3}$ ~~&~~ 0.7007 ~~&~~ 1.87 $\times~10^{-2}$ ~~&~~ -1.44$\times~10^{-2}$ \\
  2.41 ~~&~~ 0.2710 ~~&~~ -8.98 $\times~10^{-3}$ ~~&~~ 0.7137 ~~&~~ 1.74 $\times~10^{-2}$ ~~&~~ -1.48$\times~10^{-2}$ \\
  2.57 ~~&~~ 0.2713 ~~&~~ -8.24 $\times~10^{-3}$ ~~&~~ 0.7299 ~~&~~ 1.58 $\times~10^{-2}$ ~~&~~ -1.51$\times~10^{-2}$ \\
  3.06 ~~&~~ 0.2716 ~~&~~ -6.97 $\times~10^{-3}$ ~~&~~ 0.7791 ~~&~~ 1.28 $\times~10^{-2}$ ~~&~~ -1.56$\times~10^{-2}$ \\
  3.43 ~~&~~ 0.2719 ~~&~~ -6.04 $\times~10^{-3}$ ~~&~~ 0.8187 ~~&~~ 1.11 $\times~10^{-2}$ ~~&~~ -1.59$\times~10^{-2}$ \\
  3.80 ~~&~~ 0.2719 ~~&~~ -5.92 $\times~10^{-3}$ ~~&~~ 0.8419 ~~&~~ 9.26 $\times~10^{-3}$ ~~&~~ -1.53$\times~10^{-2}$ \\
  4.46 ~~&~~ 0.2723 ~~&~~ -4.79 $\times~10^{-3}$ ~~&~~ 0.9061 ~~&~~ 7.57 $\times~10^{-3}$ ~~&~~ -1.59$\times~10^{-2}$ \\
  4.76 ~~&~~ 0.2723 ~~&~~ -4.73 $\times~10^{-3}$ ~~&~~ 0.9201 ~~&~~ 6.78 $\times~10^{-3}$ ~~&~~ -1.54$\times~10^{-2}$ \\
  4.97 ~~&~~ 0.2724 ~~&~~ -4.31 $\times~10^{-3}$ ~~&~~ 0.9430 ~~&~~ 6.43 $\times~10^{-3}$ ~~&~~ -1.59$\times~10^{-2}$ \\
  5.21 ~~&~~ 0.2724 ~~&~~ -4.19 $\times~10^{-3}$ ~~&~~ 0.9531 ~~&~~ 5.95 $\times~10^{-3}$ ~~&~~ -1.60$\times~10^{-2}$ \\
\hline
\hline
\end{tabular}
\end{center}
\end{table*}

\begin{table*}[h]
\begin{center}
\caption{$(m/R)_{\infty} = 0.140$, ${\cal C}/{\cal C}_{crit} = -0.255$.}
\begin{tabular}{cccccc}
\hline
\hline
~~$d$~~&~~$M$~~&~~$BE$~~&~~$J/M_0^2$~~&~~$\Omega ~m_0$~~&~~$\Omega_s ~m_0$~~\\
\hline
  2.23 ~~&~~ 0.2704 ~~&~~ -1.11$\times~10^{-2}$ ~~&~~ 0.7273 ~~&~~ 1.92$\times~10^{-2}$ ~~&~~ -5.52$\times~10^{-3}$ \\
  2.40 ~~&~~ 0.2705 ~~&~~ -1.08$\times~10^{-2}$ ~~&~~ 0.7475 ~~&~~ 1.75$\times~10^{-2}$ ~~&~~ -5.72$\times~10^{-3}$ \\
  2.57 ~~&~~ 0.2707 ~~&~~ -1.01$\times~10^{-2}$ ~~&~~ 0.7688 ~~&~~ 1.60$\times~10^{-2}$ ~~&~~ -5.90$\times~10^{-3}$ \\
  3.26 ~~&~~ 0.2712 ~~&~~ -8.26$\times~10^{-3}$ ~~&~~ 0.8269 ~~&~~ 1.16$\times~10^{-2}$ ~~&~~ -6.69$\times~10^{-3}$ \\
  3.94 ~~&~~ 0.2716 ~~&~~ -6.97$\times~10^{-3}$ ~~&~~ 0.8913 ~~&~~ 8.92$\times~10^{-3}$ ~~&~~ -7.09$\times~10^{-3}$ \\
  4.63 ~~&~~ 0.2719 ~~&~~ -6.03$\times~10^{-3}$ ~~&~~ 0.9571 ~~&~~ 7.04$\times~10^{-3}$ ~~&~~ -7.10$\times~10^{-3}$ \\
  5.14 ~~&~~ 0.2720 ~~&~~ -5.57$\times~10^{-3}$ ~~&~~ 0.9945 ~~&~~ 6.08$\times~10^{-3}$ ~~&~~ -7.15$\times~10^{-3}$ \\
\end{tabular}
\end{center}
\end{table*}

\begin{table*}[h]
\begin{center}
\caption{$(m/R)_{\infty} = 0.140$, ${\cal C}/{\cal C}_{crit} = 0.0$.}
\begin{tabular}{cccccc}
\hline
\hline
~~$d$~~&~~$M$~~&~~$BE$~~&~~$J/M_0^2$~~&~~$\Omega ~m_0$~~&~~$\Omega_s ~m_0$~~\\
\hline
  1.89 ~~&~~ 0.2698 ~~&~~ -1.31$\times~10^{-2}$ ~~&~~ 0.7494 ~~&~~ 2.44$\times~10^{-2}$ ~~&~~ 3.29$\times~10^{-3}$ \\
  2.06 ~~&~~ 0.2701 ~~&~~ -1.21$\times~10^{-2}$ ~~&~~ 0.7689 ~~&~~ 2.19$\times~10^{-2}$ ~~&~~ 2.71$\times~10^{-3}$ \\
  2.38 ~~&~~ 0.2705 ~~&~~ -1.07$\times~10^{-2}$ ~~&~~ 0.8025 ~~&~~ 1.82$\times~10^{-2}$ ~~&~~ 1.93$\times~10^{-3}$ \\
  2.74 ~~&~~ 0.2708 ~~&~~ -9.93$\times~10^{-3}$ ~~&~~ 0.8313 ~~&~~ 1.48$\times~10^{-2}$ ~~&~~ 1.31$\times~10^{-3}$ \\
  3.49 ~~&~~ 0.2713 ~~&~~ -8.15$\times~10^{-3}$ ~~&~~ 0.8998 ~~&~~ 1.06$\times~10^{-2}$ ~~&~~ 0.73$\times~10^{-3}$ \\
  3.94 ~~&~~ 0.2715 ~~&~~ -7.36$\times~10^{-3}$ ~~&~~ 0.9421 ~~&~~ 8.92$\times~10^{-3}$ ~~&~~ 0.55$\times~10^{-3}$ \\
  4.71 ~~&~~ 0.2719 ~~&~~ -6.06$\times~10^{-3}$ ~~&~~ 1.0077 ~~&~~ 6.98$\times~10^{-3}$ ~~&~~ 0.36$\times~10^{-3}$ \\
  5.46 ~~&~~ 0.2721 ~~&~~ -5.50$\times~10^{-3}$ ~~&~~ 1.0549 ~~&~~ 5.56$\times~10^{-3}$ ~~&~~ 0.24$\times~10^{-3}$ \\
  5.83 ~~&~~ 0.2721 ~~&~~ -5.31$\times~10^{-3}$ ~~&~~ 1.0889 ~~&~~ 5.11$\times~10^{-3}$ ~~&~~ 0.22$\times~10^{-3}$ \\
\end{tabular}
\end{center}
\end{table*}

\begin{table*}[h]
\begin{center}
\caption{$(m/R)_{\infty} = 0.140$, ${\cal C}/{\cal C}_{crit} = 0.255$.}
\begin{tabular}{cccccc}
\hline
\hline
~~$d$~~&~~$M$~~&~~$BE$~~&~~$J/M_0^2$~~&~~$\Omega ~m_0$~~&~~$\Omega_s ~m_0$~~\\
\hline
  2.18 ~~&~~ 0.2703 ~~&~~ -1.14$\times~10^{-2}$ ~~&~~ 0.8198 ~~&~~ 2.02$\times~10^{-2}$ ~~&~~ 8.12$\times~10^{-3}$ \\
  2.52 ~~&~~ 0.2706 ~~&~~ -1.03$\times~10^{-2}$ ~~&~~ 0.8500 ~~&~~ 1.65$\times~10^{-2}$ ~~&~~ 7.87$\times~10^{-3}$ \\
  2.74 ~~&~~ 0.2709 ~~&~~ -9.56$\times~10^{-3}$ ~~&~~ 0.8751 ~~&~~ 1.49$\times~10^{-2}$ ~~&~~ 7.83$\times~10^{-3}$ \\
  3.26 ~~&~~ 0.2712 ~~&~~ -8.29$\times~10^{-3}$ ~~&~~ 0.9235 ~~&~~ 1.18$\times~10^{-2}$ ~~&~~ 7.51$\times~10^{-3}$ \\
  3.77 ~~&~~ 0.2715 ~~&~~ -7.31$\times~10^{-3}$ ~~&~~ 0.9703 ~~&~~ 9.69$\times~10^{-3}$ ~~&~~ 7.38$\times~10^{-3}$ \\
  4.63 ~~&~~ 0.2719 ~~&~~ -6.15$\times~10^{-3}$ ~~&~~ 1.0413 ~~&~~ 7.12$\times~10^{-3}$ ~~&~~ 7.27$\times~10^{-3}$ \\
  5.14 ~~&~~ 0.2720 ~~&~~ -5.63$\times~10^{-3}$ ~~&~~ 1.0810 ~~&~~ 6.08$\times~10^{-3}$ ~~&~~ 7.25$\times~10^{-3}$ \\
  5.66 ~~&~~ 0.2721 ~~&~~ -5.24$\times~10^{-3}$ ~~&~~ 1.1163 ~~&~~ 5.19$\times~10^{-3}$ ~~&~~ 7.19$\times~10^{-3}$ \\
\end{tabular}
\end{center}
\end{table*}

\begin{table*}[h]
\begin{center}
\caption{$(m/R)_{\infty} = 0.140$, ${\cal C}/{\cal C}_{crit} = 0.510$.}
\begin{tabular}{cccccc}
\hline
\hline
~~$d$~~&~~$M$~~&~~$BE$~~&~~$J/M_0^2$~~&~~$\Omega ~m_0$~~&~~$\Omega_s ~m_0$~~\\
\hline
  2.23 ~~&~~ 0.2707 ~~&~~ -1.01$\times~10^{-2}$ ~~&~~ 0.8716 ~~&~~ 2.03$\times~10^{-2}$ ~~&~~ 1.35$\times~10^{-2}$ \\
  2.44 ~~&~~ 0.2710 ~~&~~ -9.40$\times~10^{-3}$ ~~&~~ 0.8836 ~~&~~ 1.77$\times~10^{-2}$ ~~&~~ 1.33$\times~10^{-2}$ \\
  2.69 ~~&~~ 0.2712 ~~&~~ -8.67$\times~10^{-3}$ ~~&~~ 0.9072 ~~&~~ 1.51$\times~10^{-2}$ ~~&~~ 1.34$\times~10^{-2}$ \\
  2.77 ~~&~~ 0.2712 ~~&~~ -8.41$\times~10^{-3}$ ~~&~~ 0.9159 ~~&~~ 1.46$\times~10^{-2}$ ~~&~~ 1.35$\times~10^{-2}$ \\
  3.03 ~~&~~ 0.2715 ~~&~~ -7.63$\times~10^{-3}$ ~~&~~ 0.9452 ~~&~~ 1.27$\times~10^{-2}$ ~~&~~ 1.36$\times~10^{-2}$ \\
  3.70 ~~&~~ 0.2719 ~~&~~ -6.15$\times~10^{-3}$ ~~&~~ 1.0128 ~~&~~ 9.73$\times~10^{-3}$ ~~&~~ 1.38$\times~10^{-2}$ \\
  3.95 ~~&~~ 0.2720 ~~&~~ -5.89$\times~10^{-3}$ ~~&~~ 1.0275 ~~&~~ 8.92$\times~10^{-3}$ ~~&~~ 1.38$\times~10^{-2}$ \\
  4.79 ~~&~~ 0.2723 ~~&~~ -4.89$\times~10^{-3}$ ~~&~~ 1.0910 ~~&~~ 6.77$\times~10^{-3}$ ~~&~~ 1.41$\times~10^{-2}$ \\
  5.38 ~~&~~ 0.2725 ~~&~~ -4.09$\times~10^{-3}$ ~~&~~ 1.1563 ~~&~~ 5.71$\times~10^{-3}$ ~~&~~ 1.41$\times~10^{-2}$ \\
  5.97 ~~&~~ 0.2726 ~~&~~ -3.90$\times~10^{-3}$ ~~&~~ 1.1779 ~~&~~ 4.81$\times~10^{-3}$ ~~&~~ 1.42$\times~10^{-2}$ \\
\end{tabular}
\end{center}
\end{table*}

\begin{table*}[h]
\begin{center}
\caption{$(m/R)_{\infty} = 0.190$, ${\cal C}/{\cal C}_{crit} = -0.124$.}
\begin{tabular}{cccccc}
\hline
\hline
~~$d$~~&~~$M$~~&~~$BE$~~&~~$J/M_0^2$~~&~~$\Omega ~m_0$~~&~~$\Omega_s ~m_0$~~\\
\hline
  2.69 ~~&~~ 0.3186 ~~&~~ -1.10$\times~10^{-2}$ ~~&~~ 0.7331 ~~&~~ 1.93$\times~10^{-2}$ ~~&~~ -4.08$\times~10^{-3}$ \\
  3.10 ~~&~~ 0.3190 ~~&~~ -9.92$\times~10^{-3}$ ~~&~~ 0.7608 ~~&~~ 1.58$\times~10^{-2}$ ~~&~~ -4.66$\times~10^{-3}$ \\
  3.28 ~~&~~ 0.3192 ~~&~~ -9.34$\times~10^{-3}$ ~~&~~ 0.7838 ~~&~~ 1.48$\times~10^{-2}$ ~~&~~ -4.78$\times~10^{-3}$ \\
  3.51 ~~&~~ 0.3193 ~~&~~ -8.96$\times~10^{-3}$ ~~&~~ 0.7956 ~~&~~ 1.34$\times~10^{-2}$ ~~&~~ -4.92$\times~10^{-3}$ \\
  4.03 ~~&~~ 0.3197 ~~&~~ -7.97$\times~10^{-3}$ ~~&~~ 0.8358 ~~&~~ 1.11$\times~10^{-2}$ ~~&~~ -5.24$\times~10^{-3}$ \\
  4.29 ~~&~~ 0.3198 ~~&~~ -7.59$\times~10^{-3}$ ~~&~~ 0.8516 ~~&~~ 1.01$\times~10^{-2}$ ~~&~~ -5.39$\times~10^{-3}$ \\
  5.04 ~~&~~ 0.3202 ~~&~~ -6.61$\times~10^{-3}$ ~~&~~ 0.9089 ~~&~~ 8.07$\times~10^{-3}$ ~~&~~ -5.50$\times~10^{-3}$ \\
\end{tabular}
\end{center}
\end{table*}

\begin{table*}[h]
\begin{center}
\caption{$(m/R)_{\infty} = 0.190$, ${\cal C}/{\cal C}_{crit} = 0.0$.}
\begin{tabular}{cccccc}
\hline
\hline
~~$d$~~&~~$M$~~&~~$BE$~~&~~$J/M_0^2$~~&~~$\Omega ~m_0$~~&~~$\Omega_s ~m_0$~~\\
\hline
  2.52 ~~&~~ 0.3182 ~~&~~ -1.19$\times~10^{-2}$ ~~&~~ 0.7415 ~~&~~ 2.11$\times~10^{-2}$ ~~&~~ 2.34$\times~10^{-3}$ \\
  2.86 ~~&~~ 0.3187 ~~&~~ -1.08$\times~10^{-2}$ ~~&~~ 0.7615 ~~&~~ 1.77$\times~10^{-2}$ ~~&~~ 1.69$\times~10^{-3}$ \\
  3.03 ~~&~~ 0.3189 ~~&~~ -1.02$\times~10^{-2}$ ~~&~~ 0.7799 ~~&~~ 1.65$\times~10^{-2}$ ~~&~~ 1.52$\times~10^{-3}$ \\
  3.45 ~~&~~ 0.3192 ~~&~~ -9.34$\times~10^{-3}$ ~~&~~ 0.8036 ~~&~~ 1.36$\times~10^{-2}$ ~~&~~ 1.07$\times~10^{-3}$ \\
  3.70 ~~&~~ 0.3194 ~~&~~ -8.69$\times~10^{-3}$ ~~&~~ 0.8308 ~~&~~ 1.25$\times~10^{-2}$ ~~&~~ 0.94$\times~10^{-3}$ \\
  4.37 ~~&~~ 0.3198 ~~&~~ -7.63$\times~10^{-3}$ ~~&~~ 0.8772 ~~&~~ 9.86$\times~10^{-3}$ ~~&~~ 0.62$\times~10^{-3}$ \\
  4.96 ~~&~~ 0.3201 ~~&~~ -6.87$\times~10^{-3}$ ~~&~~ 0.9193 ~~&~~ 8.26$\times~10^{-3}$ ~~&~~ 0.46$\times~10^{-3}$ \\
  5.72 ~~&~~ 0.3202 ~~&~~ -6.42$\times~10^{-3}$ ~~&~~ 0.9481 ~~&~~ 6.58$\times~10^{-3}$ ~~&~~ 0.32$\times~10^{-3}$ \\
\end{tabular}
\end{center}
\end{table*}

\begin{table*}[h]
\begin{center}
\caption{$(m/R)_{\infty} = 0.190$, ${\cal C}/{\cal C}_{crit} = 0.247$.}
\begin{tabular}{cccccc}
\hline
\hline
~~$d$~~&~~$M$~~&~~$BE$~~&~~$J/M_0^2$~~&~~$\Omega ~m_0$~~&~~$\Omega_s ~m_0$~~\\
\hline
  2.64 ~~&~~ 0.3185 ~~&~~ -1.11$\times~10^{-2}$ ~~&~~ 0.7879 ~~&~~ 2.00$\times~10^{-2}$ ~~&~~ 1.32$\times~10^{-2}$ \\
  2.77 ~~&~~ 0.3186 ~~&~~ -1.07$\times~10^{-2}$ ~~&~~ 0.7977 ~~&~~ 1.85$\times~10^{-2}$ ~~&~~ 1.31$\times~10^{-2}$ \\
  3.12 ~~&~~ 0.3189 ~~&~~ -9.79$\times~10^{-3}$ ~~&~~ 0.8224 ~~&~~ 1.55$\times~10^{-2}$ ~~&~~ 1.26$\times~10^{-2}$ \\
  3.57 ~~&~~ 0.3193 ~~&~~ -8.67$\times~10^{-3}$ ~~&~~ 0.8642 ~~&~~ 1.29$\times~10^{-2}$ ~~&~~ 1.29$\times~10^{-2}$ \\
  4.09 ~~&~~ 0.3195 ~~&~~ -8.07$\times~10^{-3}$ ~~&~~ 0.8899 ~~&~~ 1.09$\times~10^{-2}$ ~~&~~ 1.30$\times~10^{-2}$ \\
  4.36 ~~&~~ 0.3197 ~~&~~ -7.49$\times~10^{-3}$ ~~&~~ 0.9189 ~~&~~ 9.61$\times~10^{-3}$ ~~&~~ 1.30$\times~10^{-2}$ \\
  5.80 ~~&~~ 0.3204 ~~&~~ -5.86$\times~10^{-3}$ ~~&~~ 0.9954 ~~&~~ 6.48$\times~10^{-3}$ ~~&~~ 1.20$\times~10^{-2}$ \\
\end{tabular}
\end{center}
\end{table*}

\end{appendix}



\begin{thebibliography}{99}
\bibitem{footnote3} In the inertial frame $\beta_I^i = \beta^i - (\vec{\Omega} \times \vec{r})^i$, where $\beta^i$ is given by Eq. (\ref{beta1}).
\bibitem{footnote5} For a $\Gamma=2$ polytrope, the maximum compaction ratio for a neutron star in isolation is 0.216.
\bibitem{footnote6} The approximate boundary condition for the scalar potential $\lambda$ (\ref{lambda_bc}) was based on the assumption that the stellar surface does not stray drastically from the spherical shape. The deformation from sphericity is at its worst for the closest separation distances (Fig. \ref{Cont_rho_cusp}), undermining the validity of this approximation for those cases. However, its use is still justified by the fact that the influence of the field $\lambda$ in the final result is rather small (Table \ref{Table_comp}) and
that these orbits are very likely to be inside the ISCO \cite{Marronetti:2003aa}.
\bibitem{footnote8} These orbits are acceptable representations of the binary inspiral only in the quasistationary regime.
\end{thebibliography}
\end{document}